\documentclass[10pt,aps,twocolumn,nofootinbib,tightenlines,floatfix,longbibliography]{revtex4-1}
\usepackage{amsmath,amssymb}
\usepackage{graphicx}
\usepackage{color}
\usepackage{comment}
\usepackage{bbold}
\usepackage{soul,color}
\sethlcolor{white}
\usepackage{booktabs}
\usepackage{array}
\usepackage{hyperref}
\hypersetup{
    colorlinks=true,       
    linkcolor=blue,          
    citecolor=blue,        
    filecolor=blue,      
    urlcolor=blue           
}

\def\ij{{i\dots j}}
\def\pl{\partial}
\def\mn{{\mu\nu}}

\begin{document}

\title{Hadronic transport coefficients from the linear sigma model at finite temperature}

\author{Matthew Heffernan} 
\email{heffernan@physics.mcgill.ca}
\author{Sangyong Jeon}
\email{jeon@physics.mcgill.ca}
\author{Charles Gale}
\email{gale@physics.mcgill.ca}
\affiliation{Department of Physics, McGill University, 3600 University Street, Montreal, QC, H3A 2T8, Canada}

\date{\today}

\begin{abstract}
	We investigate general frameworks for calculating transport coefficients for quasiparticle theories at finite temperature. Hadronic transport coefficients are then computed using the linear sigma model (LSM). The bulk viscosity over entropy density ($\zeta/s$) is evaluated in the relaxation time approximation (RTA) and the specific shear viscosity ($\eta/s$) and static electrical conductivity ($\sigma_{el}/T$) are both obtained in the RTA and using a functional variational approach. Results are shown for different values of the scalar-isoscalar hadron vacuum mass with in-medium masses for the interacting fields. The advantages and limitations of the LSM for studies of strongly interacting matter out of equilibrium are discussed and results are compared with others in the literature. 
	
\end{abstract}

\maketitle

%
\section{Introduction\label{sec:intro}} 
The behaviour of strongly interacting matter in extreme conditions of temperature and density is the subject of a vibrant experimental program and of numerous theoretical efforts. 
One of the major achievements of relativistic heavy-ion physics is the realization that an exotic phase of nuclear matter -- the Quark-Gluon Plasma (QGP) \cite{RevModPhys.89.035001} -- has been created in experiments performed at RHIC (the Relativistic Heavy Ion Collider, at Brookhaven National Laboratory) and at the LHC (the Large Hadron Collider, at CERN). 

A related collection of theoretical breakthroughs has shown that the dynamical evolution of this QGP is amenable to hydrodynamic modeling \cite{Gale:2013da}. Hydrodynamics is able to interpret a large body of data that reflects the collectivity of the observed particles and can even make quantitative statements about local deviations from equilibrium. The response of a quantum system to some perturbation can be characterized in different ways, including by monitoring the time it takes for the system to relax back to the equilibrium state. The relaxation time can then be related to transport parameters, which are  calculable in terms of correlation functions \cite{forster}. 
In spite of the existence of a general formalism to calculate transport parameters, obtaining them from QCD has remained challenging. This has motivated their extraction from analyses of heavy-ion phenomenology \cite{Gale:2013da,Sangaline:2015isa,Bernhard:2019bmu,Paquet:2020rxl} and from effective models of the strong interaction. 

This work reports on studies of transport coefficients using the linear sigma model (LSM). The LSM, first proposed by Gell-Mann and L\'{e}vy \cite{GellMann:1960np}, is a well-studied model of a simple hadronic system and is one of the most instructive and paradigmatic field theories.
Hadronic physics estimates have yet to agree on details of the transport parameters that characterize the strong interaction, whether those come from analyzing and interpreting data or from attempts to calculate from more-or-less first principles~\cite{Prakash:1993bt,FernandezFraile:2009mi,Noronha-Hostler:2015qmd}. Therefore, it is instructive to make predictions using the LSM. In addition, there is a need to distinguish the effects of approximations from the consequences of model-dependent assumptions. This is one of the goals pursued herein. 
We adapt and develop various general theoretical techniques in order to calculate the transport coefficients of the LSM. The shear and bulk viscosity are computed and the electrical conductivity is calculated in the LSM for the first time. 
We discuss the effect on our results of the assumed value of the vacuum mass of the $\sigma$ meson using figures that broadly span the mass of the $f_0(500)$ as defined by its large reported width \cite{Pelaez:2015qba}. 

The paper is organized as follows: Sec.~\ref{sec:framework} reviews and details the theoretical framework, established along the lines of work done in Refs. \cite{quasi,Arnold:2000dr}. It begins with a survey of the LSM with some emphasis on the thermodynamic quantities that are used as inputs into both the relaxation time approximation and the variational method. Sec.~\ref{sec:RTA} includes a brief overview of transport and shows the technique for calculating transport coefficients in the relaxation time approximation. To go beyond the limitations of the RTA, a general variational technique for massive theories with elastic and inelastic reactions is developed in Sec.~\ref{sec:variation}. Finally, in Sec.~\ref{sec:results}, we produce the results of the calculations with the different techniques and compare to the literature. Additional details of calculations are provided in the appendices.

\section{Theoretical framework \label{sec:framework}}

We begin the development of the theoretical framework with an overview of the linear sigma model. Once this is established, we describe the treatment of the effective masses and the thermodynamics of the system. 

%
\subsection{The linear sigma model} \label{sec:LSM}
The LSM is a ubiquitous relativistic field theory capable of illustrating aspects of low-energy QCD \cite{Donoghue:1992dd}.  It can incorporate mean field effects and is
thermodynamically consistent. It has been used extensively as an effective model of simple hadron dynamics and is tractable and well-studied~ \cite{Dobado:2009ek,Dobado:2012zf}. Consequently, we use it to provide parametric estimates for transport coefficients of hadronic systems, to demonstrate the effectiveness of our framework, and to highlight the quantitative effects of different approximation schemes.

The classic linear sigma model Lagrangian is 
\begin{eqnarray}
\mathcal{L} &=& \frac{1}{2}(\partial_\mu {\mathbf \Phi})^2-\frac{\lambda}{4} \left({\mathbf \Phi}^2 -f_\pi^2\right)^2.
\label{eq:linsiglag1}
\end{eqnarray}
In general, the bosonic field $\mathbf \Phi$ has $N$ components. When $N = 4$, the standard practice is to ascribe the first $N-1$ components to the pion field and define the $N^{\rm th}$ as the sigma field: $\mathbf \Phi = \left \{ \vec \pi, \sigma \right \}$. Then, the LSM model is an effective theory of soft pion dynamics owing to the isomorphism between O(4) -- the symmetry of the LSM -- and SU(2)$_{\rm L}$~$\times$~SU(2)$_{\rm R}$ -- the symmetry group for two flavours of massless quarks in QCD. At low temperatures, the O($N$) symmetry is spontaneously broken to O($N-1$) and a temperature-dependent $\sigma$ condensate $v$ appears: $\Phi_N = \sigma + v$. In the classic LSM, the condensate goes to zero at a critical temperature in a second-order phase transition \cite{FTFT}. 

The LSM can be made more realistic by explicitly breaking chiral symmetry with a pion vacuum mass. 
The Lagrangian can now be written as \cite{Scavenius:2000qd,Mocsy:2001za}
\begin{eqnarray}
\mathcal{L} &=& \frac{1}{2}(\partial_\mu \sigma)^2+\frac{1}{2}(\partial_\mu \vec{\pi})^2-\frac{\lambda}{4} \left(\sigma^2 +\vec{\pi}^2-f^2\right)^2+H\sigma \hspace{0.25in}
\label{eq:linsiglag}
\end{eqnarray}
where the vacuum expectation value $v$ of the scalar field $\sigma$ is determined by the symmetry-breaking term
\begin{eqnarray}
\lambda v (v^2-f^2) &=& H.
\end{eqnarray}

The three undetermined parameters $\lambda$, $H$, and $f^2$ are determined by the vacuum values of the pion decay constant $f_\pi$ and the pion and sigma masses 
\begin{eqnarray}
\lambda &=& \frac{m_{\sigma}^2-m_{\pi}^2}{2 f_\pi^2}\\
H &=& f_\pi m_{\pi}^2\\
f^2 &=& \frac{m_{\sigma}^2 -3 m_{\pi}^2}{m_{\sigma}^2-m_{\pi}^2}f_\pi^2.
\end{eqnarray}
In this subsection, all mass symbols represent vacuum masses -- the reason for this clarification will soon become apparent. The vacuum pion mass is chosen to be $m_{\pi} = 140$ MeV, the decay constant is $f_\pi = 93$ MeV, and in this work the vacuum sigma mass will take one of the values $m_{\sigma} = \left\{ 400, 600, 900 \right\}$ MeV.
The nature of the symmetry breaking at zero temperature drastically impacts how the chiral symmetry is restored at high temperatures.

We separate the LSM Lagrangian into kinetic and potential terms and expand the sigma field into a condensate and an excitation, $\sigma \rightarrow v + \sigma $. We call the excitation $\sigma$ as it is the true $\sigma$ meson and the condensate $v$, which is the non-vanishing vacuum expectation value of the field.

We perform our calculations in the isospin pion basis, which represents the physical pions. 
The relations between the physical pions and the Cartesian pion fields are 
\begin{eqnarray}
\pi^+ &=& \frac{1}{\sqrt{2}}\left( \pi_1 + i \pi_2 \right) \label{eq:piplus}\\
\pi^- &=& \frac{1}{\sqrt{2}}\left( \pi_1 - i \pi_2 \right)\label{eq:piminus}\\
\pi^0 &=& \pi_3\label{eq:pi0}.
\end{eqnarray}

To convert the Lagrangian to the isospin pion basis, it is simple to invert these relations. Doing so allows one to trivially rewrite the Lagrangian and read off the matrix elements. For example, in the calculation of $\mathcal{M}_{\pi^{\text{a}} \pi^{\text{b}} ; \pi^{\text{c}} \pi^{\text{d}}}$, we see that from the 4-point diagram, we get a factor $-2\lambda$. From $\pi + \pi \rightarrow \sigma \rightarrow \pi + \pi$ in the s-channel, we get a factor $\frac{4 \lambda^2 v^2}{s-m_\sigma^2}$. Thus, the 4-point pion s-channel diagram is
\begin{eqnarray}
\mathcal{M}_{\pi^{\text{a}} \pi^{\text{b}} ; \pi^{\text{c}} \pi^{\text{d}}} &=& -2\lambda + \frac{4\lambda^2 v^2}{s-m_\sigma^2}\\
&=&-2\lambda \left(\frac{s-m_\pi^2}{s-m_\sigma^2}\right).\label{eq:schannel}
\end{eqnarray}
Including other processes with appropriate Kronecker deltas produces the full matrix element,
\begin{eqnarray}
&&\mathcal{M}_{\pi^{\text{a}} \pi^{\text{b}} ; \pi^{\text{c}} \pi^{\text{d}}} \\
&&= -2\lambda \left( \frac{s-m_\pi^2}{s-m_\sigma^2} \delta_{ab}\delta_{cd} + \frac{t-m_\pi^2}{t-m_\sigma^2} \delta_{ac}\delta_{bd} + \frac{u-m_\pi^2}{u-m_\sigma^2}\delta_{ad} \delta_{bc} \right)\nonumber
\end{eqnarray}
A pole clearly arises in each channel. A consistent method of handling this pole theoretically would entail a resummation that would parametrically promote the process to higher powers of $\lambda$. However, the coupling in the LSM is larger than one and the diagrammatic expansion is not convergent. We treat the LSM as an effective theory with $\lambda$ understood as a parameter adjusted to fit $\pi \textrm{--} \pi$ scattering  \cite{FTFT}. We therefore restrict the kinematics in order to bypass the singularities and the processes to tree-level. We use the Mandelstam variables in the  limit $s, t, u \to \infty$, effectively removing the 3-point interactions \cite{quasi}. This results in the following matrix elements for the LSM:

\begin{eqnarray}
\mathcal{M}_{\sigma\sigma;\sigma\sigma} &=& -6\lambda \\
\mathcal{M}_{\pi^a\pi^a;\pi^a\pi^a} &=& -6\lambda, \hspace{0.25in}a=\{0,+,-\}\\
\mathcal{M}_{\pi^+\pi^-;\pi^+\pi^-} &=& -2\lambda\\
\mathcal{M}_{\pi^0\pi^0;\sigma\sigma} &=& -2\lambda \\
\mathcal{M}_{\pi^a \sigma;\pi^a \sigma} &=& -2\lambda, \hspace{0.25in}a=\{0,+,-\}\\
\mathcal{M}_{\pi^0\pi^b;\pi^0\pi^b} &=&-2\lambda, \hspace{0.25in}b=\{+,-\}
\end{eqnarray}

As we wish to study the critical dynamics at finite mass and temperature we adopt an approximate lower bound of $T=150$ MeV:  a common value for thermal freeze-out in studies of heavy ion collisions, which is also in the vicinity of the crossover temperature obtained in lattice calculations with baryonless QCD \cite{Bazavov:2018mes}. Higher temperatures will also be explored so that the behavior of physical quantities of interest -- such as the transport coefficients -- can be studied through gradual chiral symmetry restoration in the LSM. 

%
\subsection{Thermodynamic quantities}

To ensure consistency in the calculation, it is imperative to rigorously incorporate the thermodynamics of the LSM.
This will directly reflect chiral symmetry restoration and will be critical in determining the transport coefficients. The interactions in the LSM are evaluated in the mean field limit, which can in turn be absorbed in a mass redefinition. Detailed discussions of thermal effective masses exist in the literature, e.g. in \cite{bowman_article} (based on the methods of \cite{Carter:1996rf,PhysRevC.61.045206,PhysRevC.70.015204}); we present a brief overview of the method here for the sake of completeness. In this study, we use classical (Boltzmann) statistics
\begin{eqnarray}
	f_a^{eq}(\mathbf{p},\mathbf{x},t) = \exp(-E_a/T)
\end{eqnarray} 
which results in simplifications of statistical factors throughout.

Mean field equations of motion are derived by taking the thermal average of the Euler-Lagrange equation with respect to a general field, which we denote $\psi_a = \{\vec{\pi},\sigma\}$, 
\begin{eqnarray}
\langle \pl^2 \psi_a \rangle + \left\langle \frac{\pl U}{\pl \psi_a} \right\rangle = 0
\end{eqnarray}
where $U = \frac{\lambda}{4} \left(\sigma^2 +\vec{\pi}^2-f^2\right)^2-H\sigma$. This can be simplified further by recognizing
\begin{eqnarray}
\langle \pl^2 \psi_a \rangle = -m_a^2 \langle \psi_a \rangle = 0
\end{eqnarray}
as $\langle \psi_a \rangle = 0$. Therefore, the thermal average equation of motion becomes
\begin{eqnarray}
\left\langle \frac{\pl U}{\pl \psi_a} \right\rangle = 0.
\end{eqnarray}
Through some further calculations, it is clear that the non-trivial solution is
\begin{eqnarray}
m_a^2 = \left\langle \frac{\pl^2 U}{\pl \psi_a^2} \right\rangle. \label{eq:masstherm}
\end{eqnarray}

\begin{figure}[t!]
    \includegraphics[width=\columnwidth]{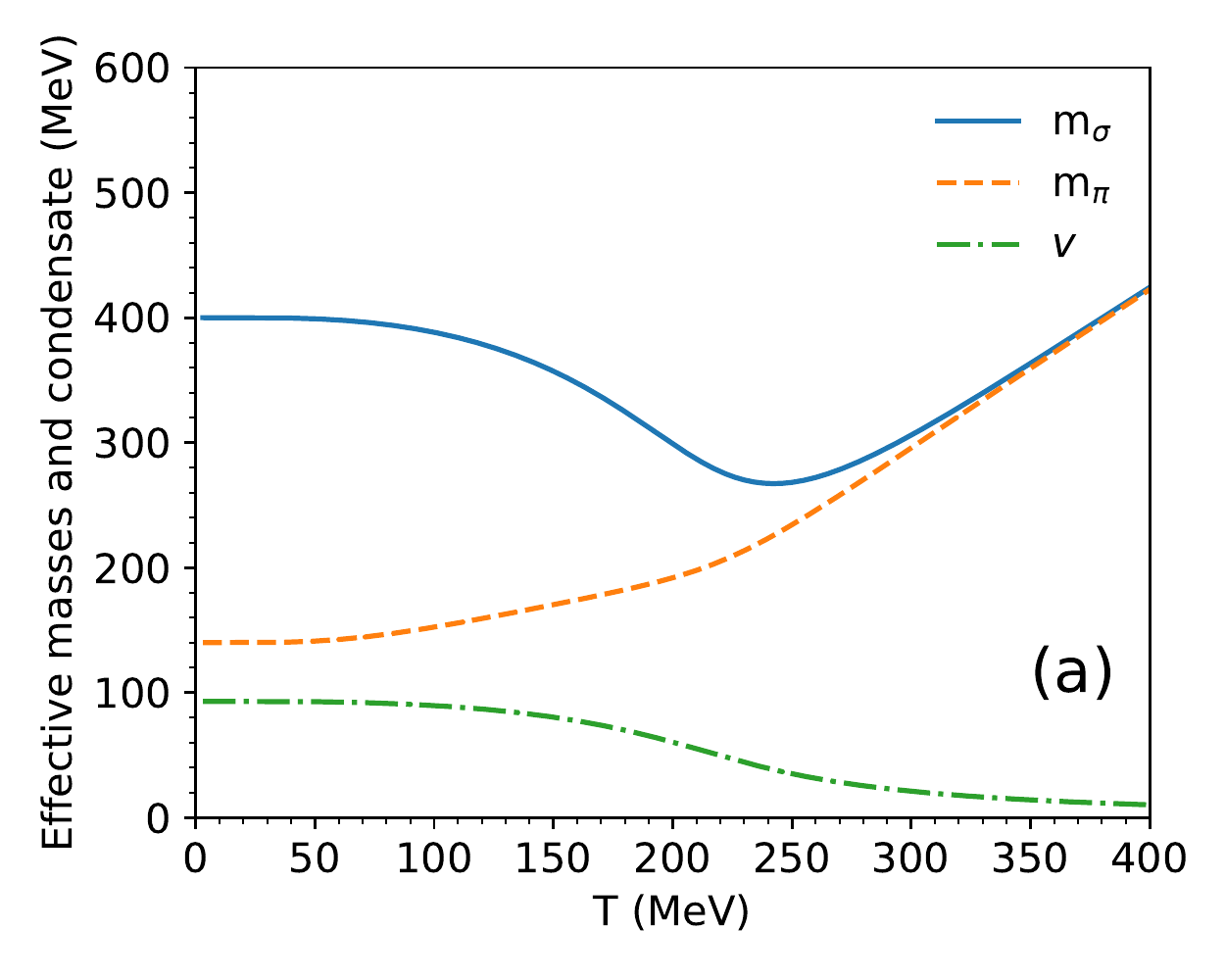}
    \newline
    \includegraphics[width=\columnwidth]{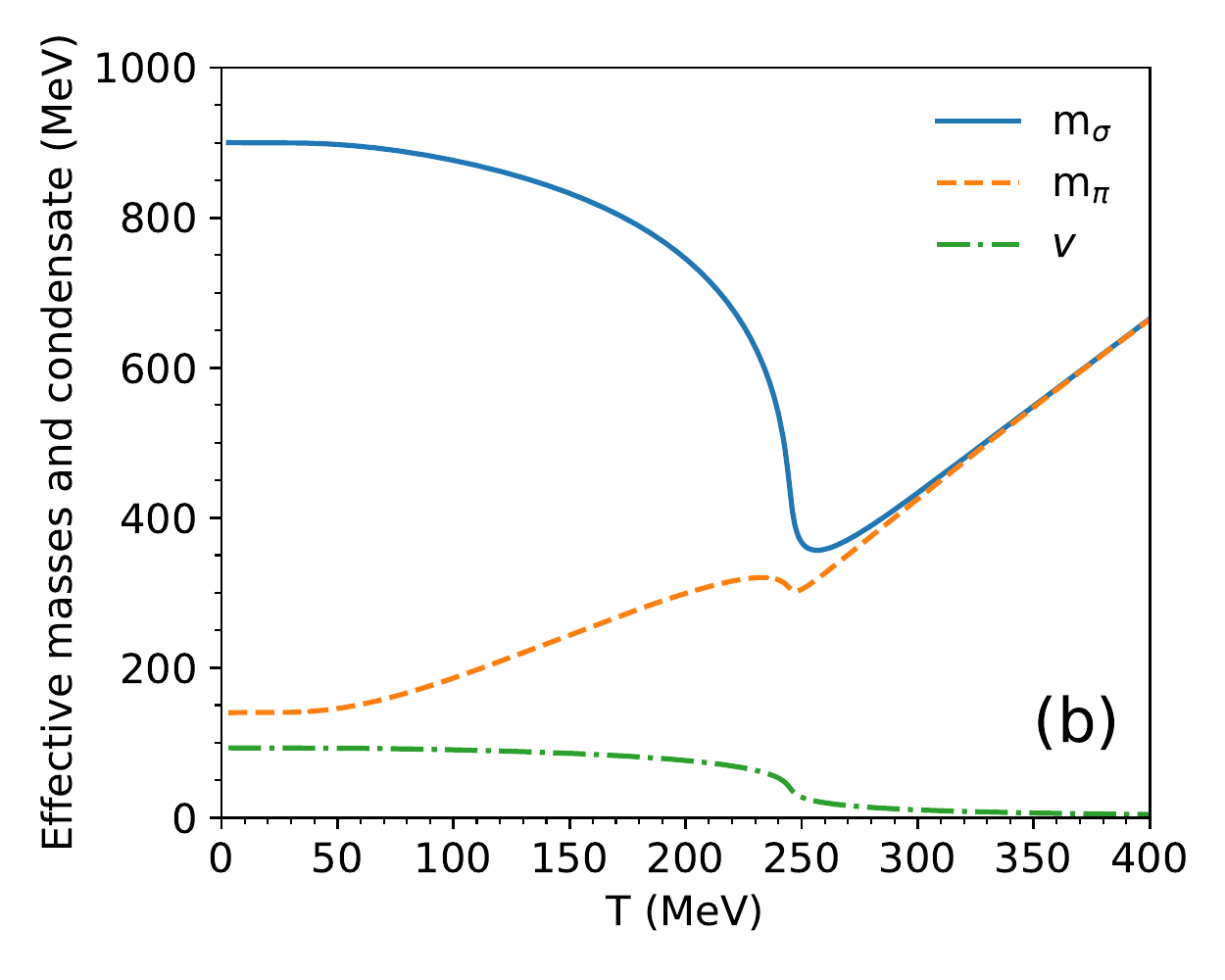}
\caption{Effective masses of the $\sigma$, $\pi$, and of the condensate for different sigma vacuum masses: (a) 400 MeV, (b) 900 MeV.} 
    \label{fig:effmass400}
\end{figure}

In order to calculate the effective masses, we solve the equation of motion for each field in the Lagrangian, producing three coupled equations that must be solved self-consistently.
These are
\begin{eqnarray}
0 &=& \lambda v \left(v^2-f^2 + 3\langle \pi^2\rangle +3 \langle \sigma^2 \rangle \right)-H\\
m_\sigma^2 &=& \lambda\left( 3v^2+3\langle\pi^2\rangle +3\langle \sigma^2\rangle -f^2 \right) \\
m_\pi^2 &=& \lambda\left( v^2+5\langle\pi^2\rangle +\langle \sigma^2\rangle -f^2 \right)
\end{eqnarray}
where the thermal average of the fields is given by 
\begin{align}\left\langle\psi_a^{2}\right\rangle=\int \frac{d^{3} p_a}{(2 \pi)^{3}} \frac{1}{E_{a}} f^{e q}_a\end{align} 
and solutions are shown in Fig.~\ref{fig:effmass400}. We show results for vacuum masses of 400 and 900 MeV, the two extreme values of the range considered here. 
While the framework we develop is general, the evaluation of scattering matrix elements, thermal effective masses, and mean-field effects is done using the LSM. Importantly, in the rest of this work all energies and masses are thermal, i.e. the single-particle energies are $E_a = \sqrt{\vec{p}_a^2+m_a^2}$, where $m$ hereon denotes the effective thermal masses discussed previously. When needed, a vacuum mass is now written as $m_{0,a}$.

As mentioned earlier, the incorporation of explicit symmetry breaking qualitatively and radically alters the symmetry restoration at high temperatures. Fig.~\ref{fig:effmass400} confirms that restoration now takes place over a broad crossover region. It is interesting to recall that our current understanding of the QCD transition from partonic to confined hadronic degrees of freedom is also a crossover (at zero net baryon density), albeit occurring at a lower temperature \cite{Bazavov:2018mes}. 

We also wish to calculate thermodynamic quantities, such as the pressure, entropy density, energy density, heat capacity, and the speed of sound. The total pressure, entropy density, energy density, and heat capacity are the sum of the contribution from each species.

The form of $T^{\mu\nu}$, discussed in more detail in Sec.~\ref{sec:boltz}, yields
\begin{eqnarray}
P &=& \frac{1}{3}T^{ii} = T \sum_a \int \frac{d^{3} p_a}{(2 \pi)^{3}} f^{eq} 
\end{eqnarray}
where we have integrated by parts, assuming a vanishing boundary term.

The energy density equation is trivial. We next provide the entropy density
\begin{eqnarray}
s &=& \frac{dP}{dT} = \frac{1}{3T^2}\sum_a \int \frac{d^{3} p_a}{(2 \pi)^{3}}|\mathbf{p}_a|^2 f^{eq}_a.
\end{eqnarray}

The calculation of the speed of sound is more involved and is shown in detail in \cite{MSc}. However, it is important to consider that the speed of sound is a thermodynamic quantity as is the consistency condition (also detailed in \cite{MSc}) and the heat capacity, Eq.~(\ref{eq:cv}). Thus, these become involved in Landau matching, which we address in Appendix~\ref{appMF}. 

Finally, the heat capacity at constant volume can be obtained by assembling expressions from Appendix~\ref{appMF}. Specifically, the combination of Eqs.~(\ref{eq:ftilde},~\ref{eq:deltaepsilon},~\ref{eq:deltatmunu}) leads directly to 
\begin{eqnarray}
c_{v} &=& \frac{d\epsilon}{dT} = \frac{1}{T^2}\sum_a \int \frac{d^{3} p_a}{(2 \pi)^{3}} \left(E_a^2 - T^2 \frac{d {m}_a^2}{dT^2} \right)f^{eq} \label{eq:cv}
\end{eqnarray}

The discussion of mean fields gains additional importance as we outline difficulties in performing exact calculations of the bulk viscosity in single and multi-component gases in Sec.~\ref{sec:variation}.

%
\section{The Relaxation Time Approximation \label{sec:RTA}}

In order to discuss transport coefficients, we begin with a discussion of transport with the Boltzmann equation (BE).
\subsection{Boltzmann equation and some definitions \label{sec:boltz}}
The Boltzmann transport equation can be written as 
\begin{eqnarray}
\frac{\pl f_a}{\pl t} + &\mathbf{v}_a& \cdot \nabla f_a = \sum_{bcd} \int_{p_b,p_c,p_d} \frac{W(a,b|c,d)}{1+\delta_{cd}} \{f_c f_d - f_af_b\}\nonumber \\
\label{eq:Boltzmann1}
\end{eqnarray}
where we have used the convenient shorthand
\begin{eqnarray}
	\int_{p} = \int \frac{d^3p}{(2\pi)^3}
\end{eqnarray}
and 
\begin{eqnarray}
	W(a,b|c,d)&=&\frac{|\mathcal{M}|^2(2 \pi)^4 \delta^4(p_a +p_b -p_c -p_d)}{16E_a E_b E_c E_d}.
\end{eqnarray}
with particles $a,b$ incoming and $c,d$ outgoing. The LSM is not explicitly restricted to $2\leftrightarrow 2$ processes, but we will only consider these processes in this work as already discussed in Sec.~\ref{sec:LSM}. 

We write the symmetric energy-momentum tensor $T^{\mu\nu}$ as
\begin{eqnarray}
	T^{\mu\nu} &=& -Pg^{\mu\nu} + w u^\mu u^\nu + \Delta T^{\mu\nu}
\end{eqnarray}
with correction terms related to the dissipation properties 
\begin{eqnarray}
	\Delta T^{\mu\nu} &=& \eta (D^\mu u^\nu + D^\nu u^\mu - \frac{2}{3} \Delta^{\mu\nu}\pl_\rho u^\rho) -\zeta \Delta^{\mu\nu} \pl_\rho u^\rho \hspace{0.2in} \label{eq:deltaTmn}
\end{eqnarray}

We choose the Landau frame, 
\begin{eqnarray}
	u_\mu \Delta T^{\mu\nu} = 0
\end{eqnarray}
where $u_\mu$ is the energy transport velocity. 
In our convention, we use the mostly negative Minkowski metric signature $(+,-,-,-)$ and define $u$ such that $u^2=1$.
In this discussion, $P$ is the pressure, $w$ is the enthalphy density $w = Ts = P+\epsilon$, $s$ is the entropy density, and $\epsilon$ is the energy density. We have defined the projection tensor and derivative normal to $u^\mu$ as 
\begin{eqnarray}
	\Delta^{\mu\nu} &=& g^{\mu\nu} - u^\mu u^\nu\\
	D_\mu &=& \pl_\mu - u_\mu u^\beta \pl_\beta
\end{eqnarray} 
respectively. We also adopt the convention that Latin indices either label species or refer to three-components of four-vectors and are thus not affected by raising and lowering. 

\subsection{Approximate solution}
The relaxation time approximation\footnote{In this paper, we will take the RTA to mean the energy-dependent relaxation time approximation.} is a popular approximation to a solution of the BE, and is commonly used in calculating transport coefficients \cite{Wiranata:2012br}. However, its validity decreases as the relaxation time increases and as such it is arguably uncontrolled. To lay the foundation for a more quantitative discussion, we begin by developing this formalism in some detail  before introducing an exact solution for transport coefficients in the LSM. We will then be able to precisely quantify this approximation in a consistent way.
We derive the forms of the shear and bulk viscosity using the Chapman-Enskog expansion \cite{chapman}.
We assume a small deviation from equilibrium; we \emph{satz an} that this is of the form
\begin{eqnarray}
	f = f^{\rm eq} + \delta f = f^{\rm eq} \left(1+\phi(p)\right) \label{eq:fexpansion}
\end{eqnarray} 
where $\phi(p)$ quantifies deviations from equilibrium. We typically suppress the momentum dependence for clarity. As a result of the maximally-general tensor decomposition of the form of Eq.~(\ref{eq:deltaTmn}), it is natural to construct $\phi$ such that it has the same decomposition
\begin{eqnarray}
	\phi_a &=& C^a_{\mu\nu} (D^\mu u^\nu + D^\nu u^\mu - \frac{2}{3} \Delta^{\mu\nu}\pl_\rho u^\rho) - A_a \Delta^{\mu\nu} \pl_\rho u^\rho \hspace{0.25in} \label{eq:phi}
\end{eqnarray}
where $C^a_{\mu\nu} = C_a p_\mu p_\nu$ and both $C_a$ and $A_a$ in general depend on the scalar $u_\alpha p^\alpha$. 

We additionally require non-decrease of entropy. In the local rest frame of the fluid, the change in the entropy is given by
\begin{eqnarray}
\partial_\mu s^\mu = \frac{\eta}{2T} \left( \partial^i u^j + \partial^j u^i + \frac{2}{3} \delta^{ij}\nabla \cdot \mathbf{u} \right)^2 + \frac{\zeta}{T}\left(\nabla \cdot \mathbf{u}\right)^2 \hspace{0.25in}
\end{eqnarray}
which requires that both shear viscosity $\eta$ and bulk viscosity $\zeta$ be non-negative. 

Assuming that interactions are localized and are point or contact interactions, the energy-momentum tensor can be written as a sum of independent contributions.
We can then return to the form $T^{\mu\nu} = T^{\mu\nu}_{\rm eq}+\Delta T^{\mu\nu}$ by writing this as 
\begin{eqnarray}
T^{\mu\nu}_{\rm eq}(x) +\Delta T^{\mu\nu}(x) &=& \sum_a \int_{p_a} \frac{p_a^\mu p_a^\nu }{E_a} f_a^{\rm eq}(x,p) \\
 &+& \sum_a \int_{p_a} \frac{p_a^\mu p_a^\nu }{E_a} f_a^{\rm eq}\phi_a(x,p).\nonumber
\end{eqnarray}
For convergence, the deviation $\phi$ must be perturbatively small, i.e. $|\phi| \ll 1$.

We now find the form of the shear and bulk viscosity in the local rest frame of the fluid. This is done by equating the two expressions for the dissipative part of the energy-momentum tensor.
\begin{eqnarray}
\Delta T^{\mu\nu}
&=& \sum_a \int_{p_a} \frac{p_a^\mu p_a^\nu }{E_a} f_a^{\rm eq}\Bigg( - A_a\partial_\rho u^\rho \label{eq:deltatmn_dep} \\
&&+C^a p_\sigma p_\gamma \left(D^\sigma u^\gamma + D^\gamma u^\sigma -\frac{2}{3} \Delta^{\gamma\sigma}\partial_\rho u^\rho \right) \Bigg)\nonumber
\end{eqnarray}

To calculate shear viscosity, we investigate a purely shear flow in a single direction. Without loss of generality, we choose $u^k = (u_x(y),0,0)$. Applying this flow to both expressions for the dissipative part of $T^\mn$ reduces them to
\begin{eqnarray}
\Delta T^{xy}(x) &=& \sum_a \int_{p_a} \frac{p_a^i p_a^j }{E_a} f_a^{\rm eq} C^a p_k p_l \left(\partial_y u_x (y) \right)
\end{eqnarray}

\begin{eqnarray}
\Delta T^{xy} &=&\eta \left(\partial_y u_x(y) \right)
\end{eqnarray}
and it is possible to identify shear viscosity as
\begin{eqnarray}
\eta&=&\frac{2}{15} \sum_a \int_{p_a} \frac{|\mathbf{p}_a|^4}{E_a} f_a^{\rm eq} C^a.
\label{shear_ChE}
\end{eqnarray}

The same method can be used to isolate the contribution of bulk viscosity that, after some manipulations, can be written as 
\begin{eqnarray}
\zeta &=& \frac{1}{3}\sum_a \int_{p_a} \frac{|\mathbf{p_a}|^2 }{E_a} f_a^{\rm eq}A_a 
\label{bulk_ChE}
\end{eqnarray}

The terms $C_a$ and $A_a$ are found by manipulations of the Boltzmann equation. 
Beginning with the LHS,
\begin{eqnarray}
\frac{\partial f_a}{\partial t} + \mathbf{v}_a \cdot \nabla f_a &=& \frac{\partial f_a}{\partial t} + \frac{\mathbf{p}_a}{E_a} \cdot \nabla f_a\\
&=& E_a^{-1} p_a^\mu \partial_\mu f_a\\
&\approx& E_a^{-1} p_a^\mu \partial_\mu f_a ^{\rm eq}.
\end{eqnarray}
The last line assumes that the off-equilibrium component $\phi_a(x,p)$ is small. 
Now we move on to the RHS. Recall Eq.~(\ref{eq:fexpansion}) and that, as a consequence of classical statistics, the product of the distribution functions before and after a collision are equivalent.
\begin{eqnarray}
f_cf_d-f_af_b 
&=& f_a^{eq}f_b^{eq} \left(\phi_d + \phi_c -\phi_a - \phi_b \right).
\end{eqnarray}

Keeping terms linear in $\phi$, we return to the Boltzmann equation.
\begin{eqnarray}
E_a^{-1} p_a^\mu \partial_\mu f_a^{\rm eq} &=& \sum_{bcd} \int_{p_b,p_c,p_d} \frac{W(a,b|c,d)}{1+\delta_{cd}}f_a^{\rm eq}f_b^{\rm eq} \label{eq:boltz}\\
&& \times \left(\phi_d + \phi_c -\phi_a - \phi_b \right) \nonumber
\end{eqnarray}
The first task is to compute the LHS. We begin with the calculation of $\partial_\mu f_a^{\rm eq}$ where $f_a^{\rm eq} = \exp(-u_\nu p^\nu / T)$ and obtain 

\begin{eqnarray}
\partial_\mu f_a^{\rm eq} &=& \partial_\mu \exp(-u_\nu p_a^\nu / T)\\
&=&-\frac{1}{T}f_a^{\rm eq}p_a^\nu \left( \partial_\mu u_\nu - \frac{1}{T}u_\nu \partial_\mu T \right) \label{eq:deriv}
\end{eqnarray}

We now rewrite the Boltzmann equation using some thermodynamic quantities. These are standard, but a detailed treatment can be found in \cite{MSc} where many standard results are collected.
Using the speed of sound, we may now make some progress in rewriting the Boltzmann equation into a more convenient form
\begin{eqnarray}
p_a^\mu \partial_\mu f_a^{eq}
&=&- \frac{1}{T}f_a^{eq}p_a^\mu p_a^\nu \big( \partial_\mu u_\nu \\
&&\hspace{0.25in}- u_\nu (u^\alpha \partial_\alpha u_\mu -v_s^2 u_\mu \partial_\alpha u^\alpha) \big). \nonumber
\end{eqnarray}

At this stage it is necessary to substitute the structure of $\phi$ and group terms in Eq.~(\ref{eq:boltz}), which now reads
\begin{eqnarray}
0 &=&\frac{f_a^{eq}p_a^\mu p_a^\nu}{2TE_a} \Big( D_\mu u_\nu + D_\nu u_\mu + \frac{2}{3}\Delta_\mn \partial_\rho u^\rho \label{eq:rewrite1}\\
&& \hspace{0.75in}- \frac{2}{3}\Delta_\mn \partial_\rho u^\rho + 2v_s^2 u_\nu u_\mu \partial_\rho u^\rho\Big) \nonumber\\ 
&+&\sum_{bcd} \int_{p_b,p_c,p_d} \frac{W(a,b|c,d)}{1+\delta_{cd}}f_a^{eq}f_b^{eq} \left(\phi_d + \phi_c -\phi_a - \phi_b \right) \nonumber.
\end{eqnarray}

To do this in a consistent way, it is necessary to consider this equation term-by-term. It is useful to define the shorthand
\begin{eqnarray}
\mathcal{D}^\mn &=& D^\mu u^\nu + D^\nu u^\mu -\frac{2}{3} \Delta^{\mu\nu}\partial_\rho u^\rho\\
\mathcal{U} &=& \partial_\rho u^\rho.
\end{eqnarray}
Rewriting the $\phi$ coefficients of each term, we find that
\begin{eqnarray}
\phi_d + \phi_c -\phi_a - \phi_b &=& - \left(A_d + A_c - A_a -A_b \right) \mathcal{U} \label{eq:phiregroup}\\
&&+ (C^d_\mn + C^c_\mn - C^a_\mn - C^b_\mn)\mathcal{D}^\mn. \nonumber
\end{eqnarray}

In taking the relaxation time approximation, we suppose that all particles are in equilibrium except for species $a$, which is out of equilibrium by a perturbatively small amount, $f_a = f_a^{\rm eq} + \delta f_a$. To that order,
\begin{eqnarray}
\frac{\partial f_a}{\partial t} + \mathbf{v}_a \cdot \nabla f_a &=& \sum_{bcd} \int_{p_b,p_c,p_d} \frac{W(a,b|c,d)}{1+\delta_{cd}}\{f_cf_d-f_af_b\}\nonumber \\
&= & - \omega_a \delta f_a
\end{eqnarray}
where
\begin{eqnarray}
\omega_a &=& \sum_{bcd} \frac{1}{1+\delta_{cd}}\int_{p_b,p_c,p_d} W(a,b|c,d)f_b^{eq} \label{eq:relaxtime}
\end{eqnarray}
is the interaction frequency. The expressions and phase space for $\omega_a$ are shown in detail in \cite{MSc}. We define the relaxation time to be 
\begin{eqnarray}
\tau_a = \omega_a^{-1}
\label{relax}
\end{eqnarray}
and the deviation $\delta f_a$ is that in Eq.~(\ref{eq:fexpansion}). Both the interaction frequency and relaxation time are energy-dependent, but we suppress this in the notation for clarity.
To find the viscosities, it is necessary to substitute the deviation $\phi$
into the Boltzmann equation. Thus, we find
\begin{eqnarray}
 \omega_a f_a^{eq}\phi_a &=& \frac{f_a^{eq}p_a^\mu p_a^\nu}{2TE_a} \Big( D_\mu u_\nu + D_\nu u_\mu + \frac{2}{3}\Delta_\mn \partial_\rho u^\rho \hspace{0.1in} \\
 && - \frac{2}{3}\Delta_\mn \partial_\rho u^\rho + 2v_s^2 u_\nu u_\mu \partial_\rho u^\rho\Big) \nonumber \hspace{0.2in}\\
&=& \omega_a f_a^{eq}\left[-A_a\partial_\rho u^\rho +C^a_\mn \mathcal{D}^\mn \right].
\end{eqnarray}
Therefore, we arrive at
\begin{eqnarray}
-A_a\mathcal{U} +C^a_\mn \mathcal{D}^\mn &=&\frac{\tau_a p_a^\mu p_a^\nu}{2TE_a} \Big( \mathcal{D}^\mn + \frac{2}{3}\Delta_\mn \mathcal{U} \\
&& \hspace{0.5in}+ 2v_s^2 u_\nu u_\mu\mathcal{U}\big)\nonumber .
\end{eqnarray}
It is then possible to make the following identifications
\begin{eqnarray}
C_a &=& \frac{\tau_a}{2TE_a}\\
A_a &=&-\frac{\tau_a}{2TE_a} \left( \frac{2}{3}p_a^\mu p_a^\nu\Delta_\mn + 2v_s^2p_a^\mu p_a^\nu u_\nu u_\mu\right)\\
&=& \frac{\tau_a}{3TE_a} \left(\left(1- 3v_s^2\right) E_a^2 - m_a^2 \right)
\end{eqnarray}
 
The shear viscosity is now readily calculated using Eq. (\ref{shear_ChE})
\begin{eqnarray}
\eta &=&\frac{1}{15T} \sum_a \int_{p_a} \frac{|\mathbf{p}_a|^4}{E_a^2} f_a^{\rm eq} \tau_a 
\end{eqnarray}

The inclusion of mean-field effects and ensuring thermodynamic consistency makes the evaluation of the bulk viscosity slightly more complicated than that of the other transport coefficients discussed in this work. This is addressed in Appendix~\ref{appMF} with details in \cite{MSc}. The result for the bulk viscosity is
\begin{eqnarray}
\zeta = \frac{1}{9T} \sum_a \int_{p_a} \frac{\tau_a}{E_a^2}f_a^{eq}\left( |\mathbf{p}_a|^2 - 3v_s^2 \left[E_a^2 - T^2 \frac{d m_a^2}{dT^2} \right]\right)^2 \nonumber \\
\label{eq:bulkmeanfield}
\end{eqnarray}
With the the incorporation of mean field effects, the bulk viscosity meets the Landau matching condition. 
We also display the result for the electrical conductivity in the RTA \cite{cercignani2002relativistic,Puglisi:2014sha,Harutyunyan:2016rxm,Dash:2020vxk} for later use: 
\begin{equation}
\sigma_{\rm el}=\frac{1}{3 T} \sum_{a} q_{a}^{2} \int_{p_a} \frac{p_a^{2}}{E_{a}^{2}} \tau_{a} f_{a}^{eq}.
\end{equation}
The only modification of $\eta$ and of the electrical conductivity $\sigma_{\rm el}$ due to mean fields comes from the presence of effective masses in phase space considerations. Finally, $\tau_a$ is calculated numerically using Eq.~(\ref{relax}) and inserted into these workings. 

\section{Variational Method \label{sec:variation}}
It is difficult to calculate systematic corrections to the RTA. As it is important to separate the impact of approximations from those of the transport properties of the medium itself, we turn to a method for calculating transport coefficients exactly in the limit of the linearized Boltzmann equation.
We extend the variational method of \cite{Arnold:2000dr,Arnold:2003zc,Arnold:2006fz} to massive theories and remove the small momentum transfer approximation so the technique is applicable to inelastic processes. We will follow the same notation and provide an overview of the method for completeness. For details of incorporating masses into this framework, see Appendix~\ref{appFuncMin}.

We begin by laying out some notation and motivations of the general form. To do this, we define a collision operator that is already linear in the deviation from equilibrium when using Boltzmann statistics.
\begin{align}
\left(\mathcal{C} \delta f\right)^{a}(p_a) \equiv & \frac{1}{1+\delta_{ab}} \sum_{b c d} \int_{p_b,p_c,p_d} W(a,b|c,d) \\
& \times f_{eq}^{a}(p_a) f_{eq}^{b}(p_b)\left[\delta f^{a}+\delta f^{b}-\delta f^{c}-\delta f^{d}\right].\nonumber
\end{align}

It may then be shown that the Boltzmann equation, to first order in gradients, is a linear integro-differential equation,
\begin{align}
\left[\partial_t+\hat{p_a} \cdot \partial_\mathbf{x}+\mathbf{F}_{\mathrm{ext}}^{a} \cdot \partial_{p_a}\right] f_{eq}^{a}(p_a, \mathbf{x}, t)=-\left(\mathcal{C} \delta f\right)^{a}(p_a, \mathbf{x}, t)
\end{align}
We now move on to interpreting the LHS. In the local fluid rest frame it may be written as 
\begin{align}
    \mathrm{LHS}=\beta f_{eq}^{a}(p_a, x) q^{a} I_{i \cdots j}(\hat{\mathbf{p}}) X_{i \cdots j}(x)
\end{align}
where $q^a$ is the conserved charge of the quantity of interest and we have separated the angular and spatial dependence into $I_\ij$, the unique rotationally covariant tensor, and $X_\ij$, the spatial tensor denoting the driving field:
\begin{align}
X_{i \cdots j}(x) \equiv\left\{\begin{array}{ll}
\nabla \cdot \mathbf{u}, & l=0\\
-E_{i}, & l=1 \\
\frac{1}{\sqrt{6}}\left(\nabla_{i} u_{j}+\nabla_{j} u_{i}-\frac{2}{3} \delta_{i j} \nabla \cdot \mathbf{u}\right), & l=2
\end{array}\right.
\end{align}
and
\begin{eqnarray}
	I_\ij(\mathbf{\hat{p}}) &\equiv& 
	\begin{cases}
	\delta_{ij} & l=0 \text{ (bulk viscosity)} \\
	\hat{p}_i & l=1 \text{ (conductivity)} \\
	\sqrt{\frac{3}{2}}(\hat{p}_i \hat{p}_j-\frac{1}{3}\delta_{ij}) & l=2 \text{ (shear viscosity)} 
	\end{cases}\hspace{0.2in}\vspace{0.1in}
\end{eqnarray}
Due to the rotational invariance of the collision operator $\mathcal{C}$, the departure from equilibrium and the driving field must have the same angular form. Thus, the deviation that will solve the Boltzmann equation must be
\begin{align}
    \delta f^{a}(p_a, x)=\beta^{2} X_{i \cdots j}(x) I_{i \cdots j}(\hat{p_a}) \chi^{a}(|p_a|).
\end{align}
In order to solve this, we additionally define an inner product as
\begin{eqnarray}
	\big(f,g\big) &=& \beta^3 \sum_a \int_{p_a} f(p_a)g(p_a)f_a^{eq}.
\end{eqnarray}
that will allow us to construct a functional, $Q$. 
We will expand this functional in a variational basis and maximize the expanded functional to determine variational coefficients and extract transport coefficients. 

We define a functional $Q[\chi_\ij]$ such that it is extremal when $\chi^a(p)$ satisfy the linear Boltzmann equation 
\begin{eqnarray}
	Q[\chi_\ij] &=& (\chi_\ij,S_\ij) - \frac{1}{2}(\chi_\ij,\mathcal{C}\chi_\ij).\label{eq:Q}
\end{eqnarray}
In the above,
\begin{eqnarray}
	&&\chi^a_\ij(p_a) = I_\ij(\hat{p}) \chi^a(p)\\
	&&S^a_\ij = -Tq^a f_0^a I_\ij(p_a)\label{eq:source}
\end{eqnarray}
and $\mathcal{C}$ is the linearized collision operator while $\chi^a(p)$ is a rotationally invariant function depending only on excitation energy.
$\chi^a(p)$ is what we will expand in a convergent variational basis in order to calculate transport coefficients.

The source in Eq.~(\ref{eq:source}) can be written in terms of an expansion basis $\phi^m$ (see Appendix~\ref{appFuncMin}) as
\begin{eqnarray}
\tilde{S}_{m} = (S_i,\phi^m) &=& -\beta^2 \sum_a \int_{p_a} f_a^{eq}q^a\phi^a_m.
\end{eqnarray}
and the collision term may be written
\begin{eqnarray}
\tilde{C}_{mn} &=& (\phi^m_i,\mathcal{C}\phi^n_i).
\end{eqnarray}
This may be assembled into the maximized functional $Q_{max}$
\begin{eqnarray}
Q_{max} &=& \frac{1}{2} \tilde{S}^T \tilde{C} \tilde{S}.
\end{eqnarray}
From this maximized functional, we are able to calculate the transport coefficients.
Keeping in mind the angular momentum structure of Eq.~(\ref{eq:source}), the expressions for transport coefficients are
\begin{eqnarray}
\eta = \frac{2}{15} Q_{max}\\
\sigma_{\rm el} = \frac{2}{3} Q_{max}\\
\zeta = 2Q_{max}
\end{eqnarray}
where details of constructing the different components are given in Appendix~\ref{appFuncMin}.
While we only calculate those applicable to our theory, the work in \cite{Arnold:2000dr} also calculates the flavor diffusion constants; our extended framework can also calculate this for massive theories with inelastic processes.

In the bulk viscosity calculation, it is necessary to orthogonalize to the zero modes. This is because the bulk viscosity is a spin 0 quantity (a scalar), while the other transport coefficients are spin 1 (a vector) or spin 2 (a tensor). Exact zero modes are only present in the scalar quantity because when taking the dot product of the momenta, angular factors in the spin 1 or 2 modes break the degeneracy.

A zero mode is an eigenvector with a vanishing eigenvalue that presents a problem to the inversion of the matrix. It corresponds to conserved quantities in the system, so in a system with only number-conserving processes, two zero modes exist: one corresponding to the conservation of energy and the other to the conservation of total particle number.
At the order we consider, our theory exactly conserves particle number, which means that the source must be absolutely orthogonal to both the energy and number conservation zero modes. This can only be accomplished by detailed accounting for a chemical potential that we do not develop, or by explicitly knowing the form of the zero mode. The exact form of the zero mode is known for a single-component gas \cite{Arnold:2006fz} but this result is not applicable to multi-component gases and rigorous study of this is beyond the scope of this work. To avoid detailed treatment of a chemical potential, one would have to make the system particle number non-conserving. One would then need to consider $1\leftrightarrow 2$ processes or $2\leftrightarrow 4$ processes, which would mean the return of poles in the matrix elements, poorly-defined expansion to higher orders in the coupling, and/or a re-evaluation of the approximations in this work. Due to a large coupling constant in the LSM, higher-order expansions are not well-defined. As a result, these considerations are beyond the scope of this work and the development of techniques to address them will be pursued elsewhere. 

The details of the LSM itself are incorporated in the thermal masses, collision term, and source of the variational framework that has been developed in this section. As a result, this method remains completely general and quantitative comparisons can be made to other general methods and calculations, such as the RTA and perturbative QCD. 

\section{Results \label{sec:results}}
We present the numerical results beginning with the thermodynamics of the LSM. This reveals that the behaviour near chiral symmetry restoration contains interesting physics that can be explored in more physical models. Importantly, the features of the chiral symmetry restoration have a consequence upon the system's thermodynamics. Having verified that we are able to resolve the dynamics we expect in the thermodynamic quantities, we compute transport coefficients beginning with the vector and tensor quantities $\sigma/T$ and $\eta/s$ in both the relaxation time approximation and the variational method as these do not possess exact zero modes. We conclude by calculating $\zeta/s$ in the relaxation time approximation, leaving the treatment of zero modes for future work.

Integrals are evaluated using Vegas adaptive Monte Carlo \cite{Vegas} and numerical uncertainties are propagated through nested calculations\footnote{\url{https://pypi.org/project/uncertainties/}}. We compare to hadron gas calculations, chiral perturbation theory, and pQCD calculations and show that the LSM demonstrates key features of these other approaches. We also use this calculation to provide insight to the possible parameters of a sigma meson, keeping in mind the characteristics of the $f_0(500)$ \cite{PDG} and also the possible caveats associated with identifying this $\sigma$ field with a physical particle \cite{Lin:1990cx}. 

\subsection{Thermodynamic quantities}

\begin{figure}[!tbp]
	\includegraphics[width=\columnwidth]{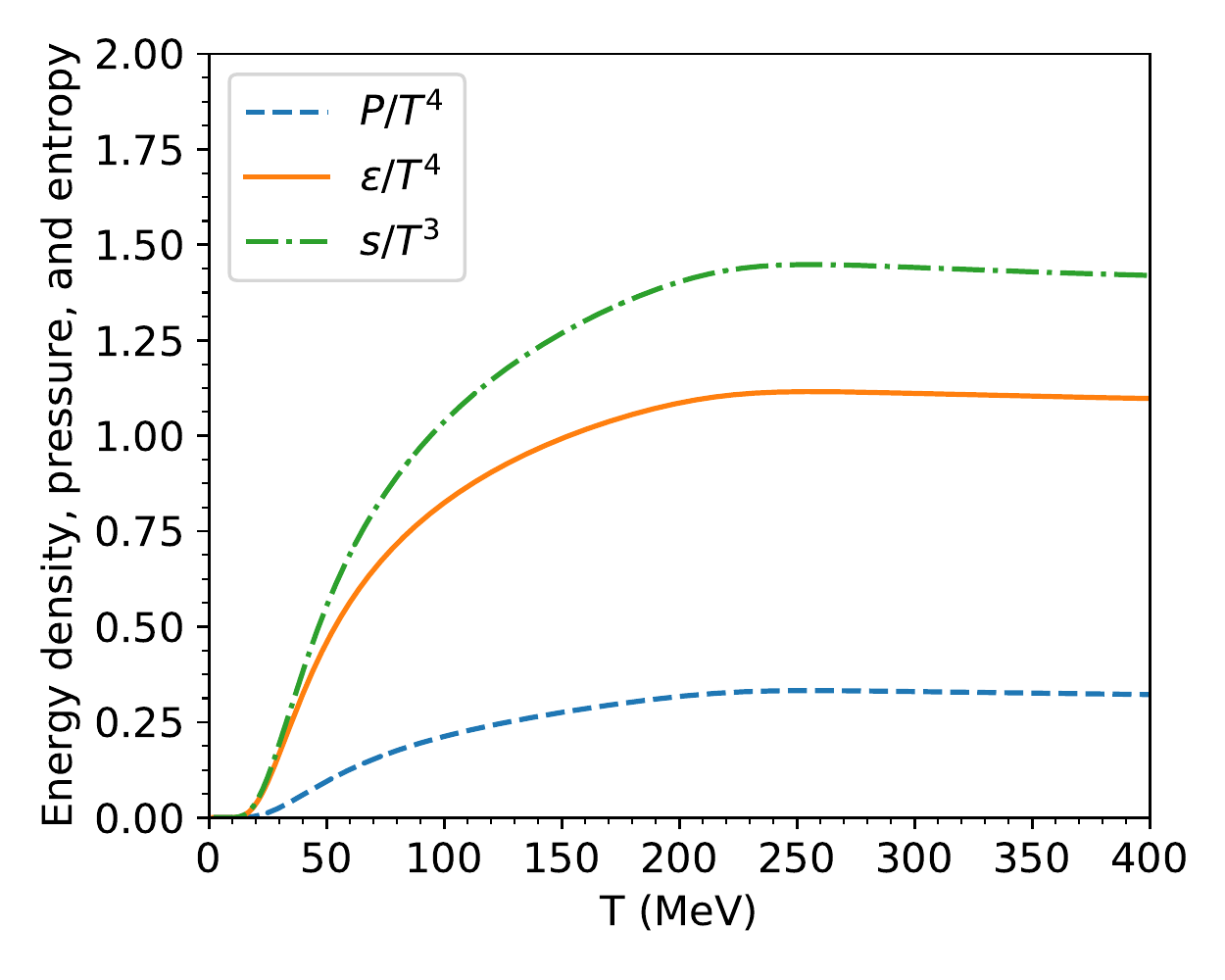}
	\caption{\label{fig:thermo400} Thermodynamic quantities for a vacuum sigma mass of 400 MeV. Quantities for other values of the vacuum sigma mass can be seen in \cite{quasi}. }
\end{figure}

\begin{figure}[!tbp]
	\includegraphics[width=\columnwidth]{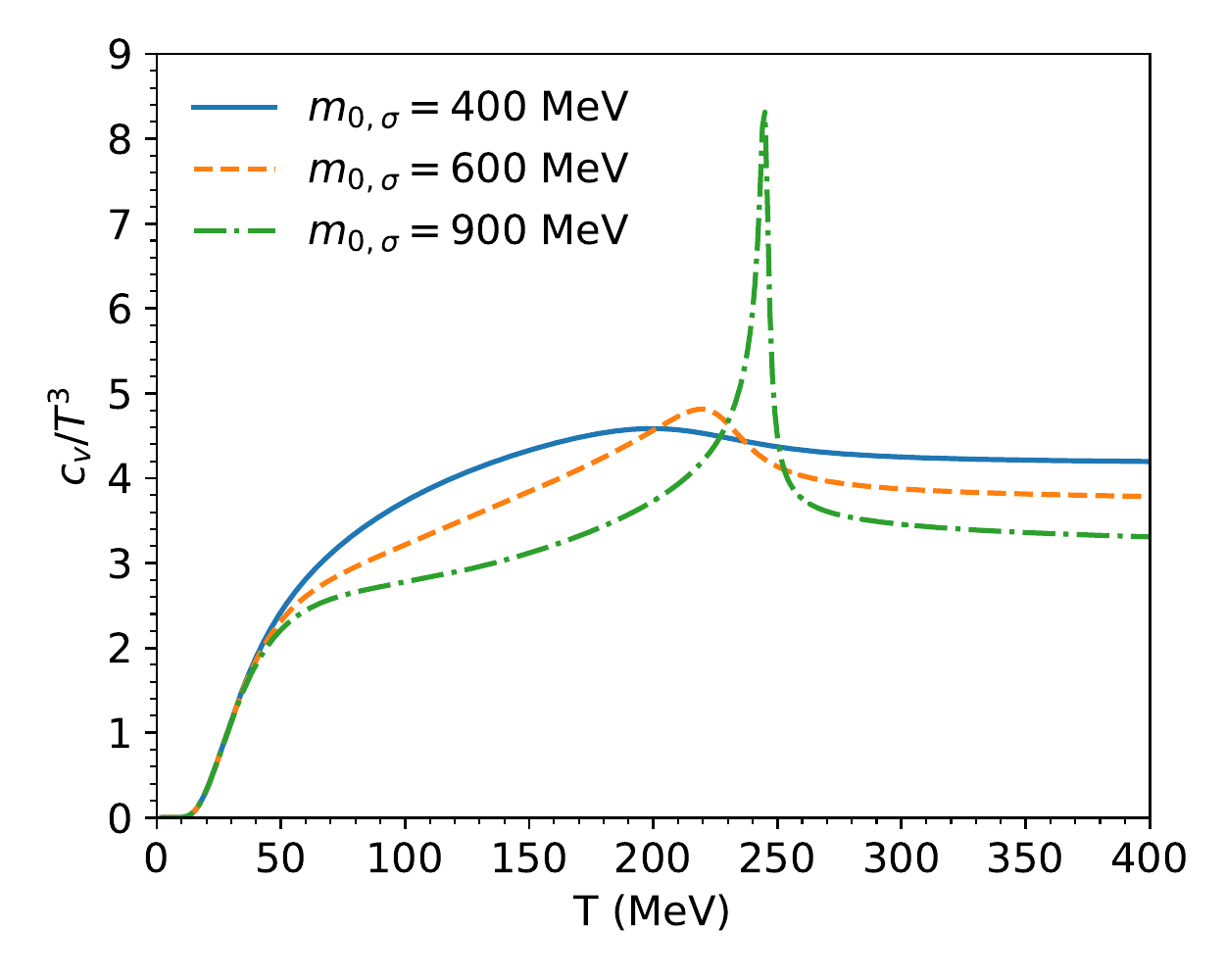}
	\caption{\label{fig:cv} Heat capacity at various values of the vacuum sigma meson mass. The 600 MeV and 900 MeV vacuum sigma mass cases match those of \cite{quasi}. }
\end{figure}

\begin{figure}[!tbp]
	\includegraphics[width=\columnwidth]{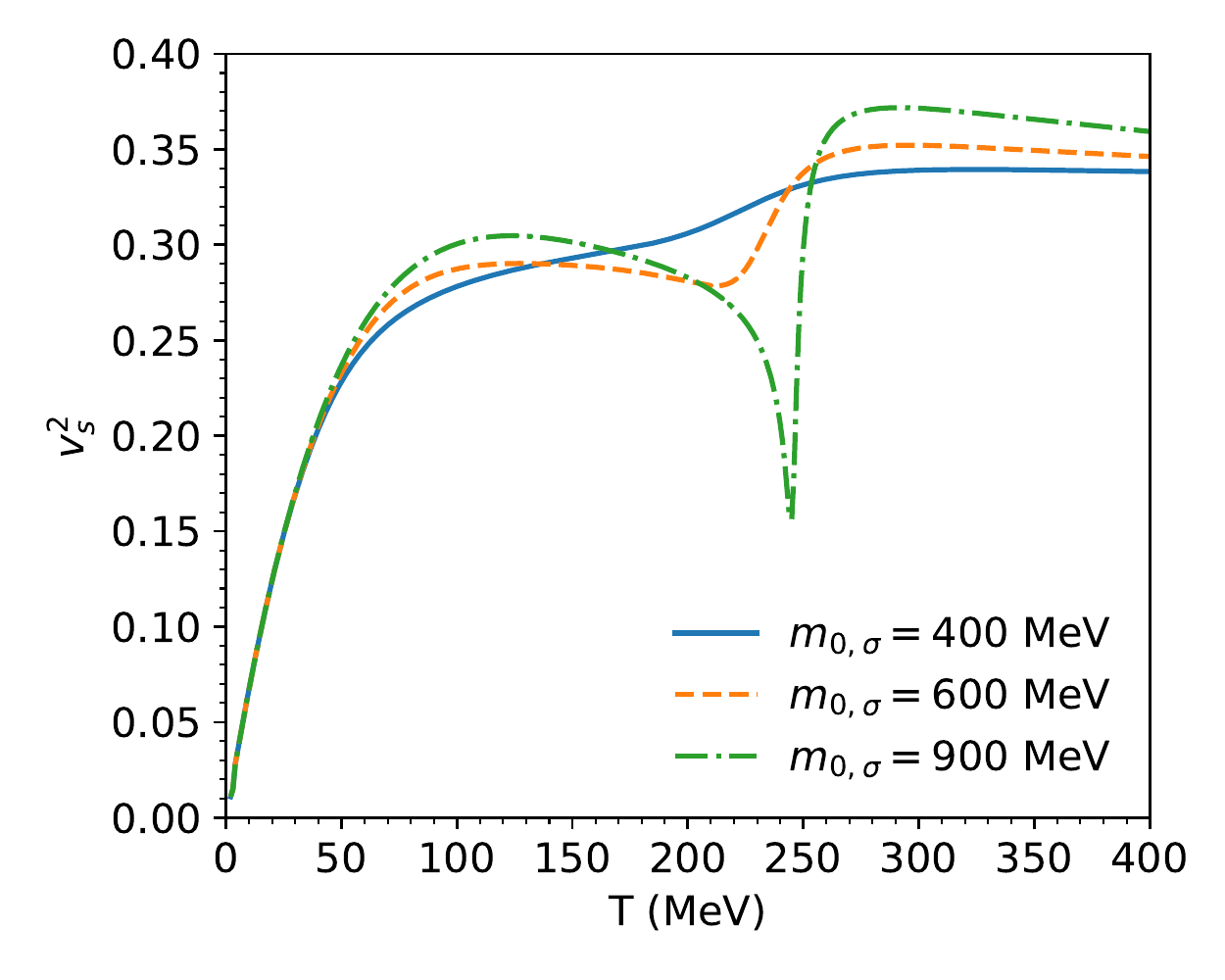}
	\caption{\label{fig:vscomparison} Speed of sound in the hadron gas. The 600 MeV and 900 MeV vacuum sigma mass cases match those of \cite{quasi}.}
\end{figure}

Once the effective masses shown in Fig.~\ref{fig:effmass400} are obtained via a self-consistent numerical optimization \cite{scipy}, one is able to calculate thermodynamic quantities such as energy density, pressure, and entropy density (Fig.~\ref{fig:thermo400}); heat capacity (Fig.~\ref{fig:cv}); and the speed of sound (Fig.~\ref{fig:vscomparison}). An important and clearly visible feature is the different behaviors of the thermodynamic quantities and heat capacity at different values of the vacuum sigma mass: the higher the vacuum sigma mass, the more suddenly chiral symmetry is restored, producing a peak in the heat capacity and a corresponding trough in the speed of sound. Similarly, the thermodynamic quantities have a more pronounced behavior at $m_{0, \sigma} = 900$ MeV than they do at 600 MeV. This behavior has a clear impact on the transport coefficients, but none more so than the bulk viscosity, which is highly sensitive to the conformality of the system as measured by the speed of sound.

\subsection{Electrical conductivity}
We now turn our attention to the transport coefficients and begin with the electrical conductivity, which quantifies the conduction properties of the medium. The DC conductivity is the real, static part of the complex conductivity tensor, $\sigma_{\rm el} = \rm{Re}\{\lim_{\omega \to 0^+} \sigma_{i i} (\omega, 0)\}$, and is related to the electrical field through Ohm's law: $\vec J_{\rm EM} = \sigma_{\rm el} \vec E$. One may use linear response theory \cite{FTFT} to derive a Kubo formula for the conductivity tensor, which features the electromagnetic current operator:
\begin{eqnarray}
\sigma_{i j} = -i \int d^4 x \theta(t) e^{i (\omega t - \vec k \cdot \vec x)} \langle \left[J_i^{\rm EM} (t, \vec x), J_j^{\rm EM} (0,0)\right]\rangle \nonumber \\
\end{eqnarray}
Importantly, the emission of electromagnetic radiation is also regulated by the current-current correlator \cite{Gale:1990pn}. Therefore, in addition to its intrinsic interest from the point of view of transport, $\sigma_{\rm el}$ can provide information on the ability of the hot and strongly interacting medium to emit soft photons:
\begin{eqnarray}
\lim_{\omega\to 0^+} \lim_{k\to 0^+} \omega \frac{d^3 R}{d^3 k} =\# T \sigma_{\rm el}
\end{eqnarray}
where \# is a numerical pre-factor. Rigorous quantitative control of the electrical conductivity can thereby also provide constraints on the phenomenology of soft electromagnetic radiation. 

The electrical conductivity has been the subject of previous studies with both hadronic (confined) and partonic degrees of freedom \cite{Arnold:2000dr,FernandezFraile:2005ka,Gagnon:2006hi,Kadam_2018,phsd,njl-dqpm,Sahoo:2018dxn,GUPTA200457,Lee:2014pwa,Borsanyi:2010cj}. The DC electrical conductivity was not previously studied in the LSM, although its vector structure is conducive to calculation using the variational method we have extended. We provide calculations here both in the relaxation time approximation and in the variational method, noting the differences between them.

Some extractions of the electrical conductivity of hadron gases are available, making this an ideal choice for further study and for validation of our method. An additional benefit comes in computational efficiency: since the conserved charge for electrical conductivity is simply the electrical charge of particle $a$, the source (Eq.~(\ref{eq:source})) for neutral particles is identically 0. This naturally simplifies the structure of the maximization, since these components will not contribute and do not have to be calculated in either the exact variational or RTA approaches. The $\sigma$ meson contributes resistance only through interactions with $\pi$. 

\begin{figure}[!htbp]
	\includegraphics[width=\columnwidth]{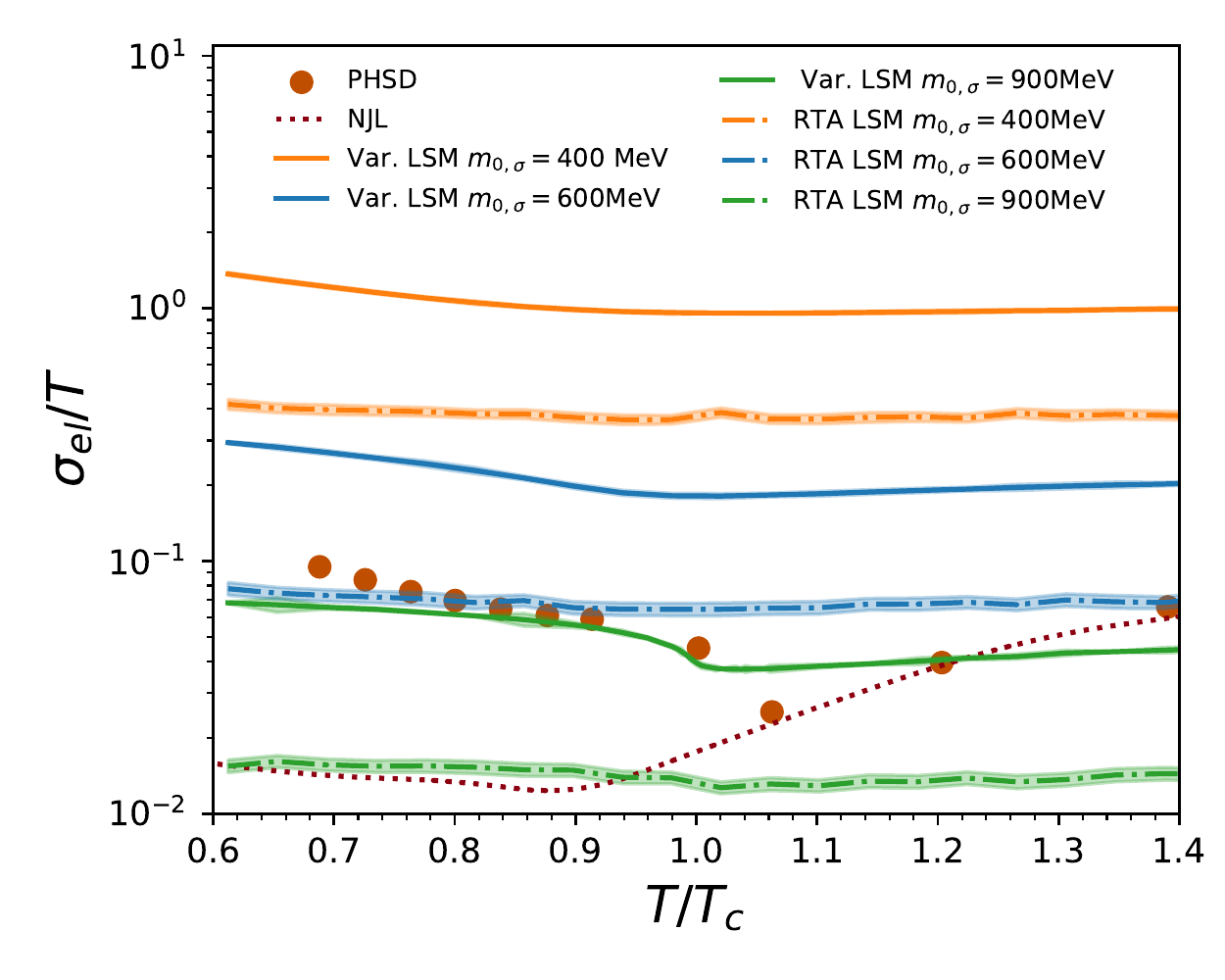}
	\caption{\label{fig:sigmacomparison} Comparison with some other calculations of electrical conductivity, for different values of the sigma vacuum mass. ``Variational" is abbreviated to ``Var." Also shown here are the electrical conductivity obtained in the Parton-Hadron String Dynamics calculation \cite{phsd} and the Nambu-Jona-Lasinio model \cite{njl-dqpm}. }
	\label{elect_cond}
\end{figure}

As current calculations of transport parameters of strongly interacting matter have not converged to a set of well-defined values, 
it is prudent to learn from this apparent lack of unity and to compare with various approaches and models. We begin with the electrical conductivity here.
Fig.~\ref{fig:sigmacomparison} contains results for the electrical conductivity over temperature plotted as a function of temperature, calculated by both the RTA and variational techniques. Some previous calculations of $\sigma_{\rm el}$ for a hadron gas are also shown. For ease of comparison between different models and results we use a scaled temperature. Since there is no genuine phase transition occurring with increasing temperature, one must adopt an operational definition of critical temperature  ``$T_c$''. Possible choices are the temperature where effective masses are minimized (see Fig. \ref{fig:effmass400}), or when $c_V/T^3$ peaks (see Fig. \ref{fig:cv}). In addition, these answers would vary, depending on the choice of $m_{0, \sigma}$. Choosing the first criterion leads to $T_c$ = (242, 245, 259) MeV, for $m_{0, \sigma}$ = (400, 600, 900) MeV, respectively. The second yields $T_c$ = (200, 219, 245) MeV. We will choose the first scheme and the intermediate value of the scalar-isoscalar mass. Therefore, for calculations that involve the LSM, we set $T_c$ = 245 MeV. The ``critical temperature'' in other approaches shown here is the  one reported using each of those models. 

Our results depend strongly on the value of the vacuum $\sigma$ mass, as seen in Fig. \ref{elect_cond}. For comparison, we also show results obtained with the PHSD \cite{phsd} and NJL \cite{njl-dqpm} models. As is the case for the magnitude of the transport parameters, there is currently no consensus on the details of their temperature dependence. A hint of critical behavior is observed in the results with large $\sigma$ mass, which is also seen in the results of PHSD, but that is where similarities end. This spread in theoretical results is also seen in more extensive compilations \cite{Greif:2018ubz}, which makes it difficult to single-out a preferred value of $m_{0, \sigma}$ by comparing between theoretical results. However, the magnitude of the vast majority of results obtained in the field fall within the range spanned by the value of $m_{0, \sigma}$ explored here. 

We directly compare the calculations in the relaxation time approximation and the variational method and we find that the two are within a factor $\sim$ 3 of each other. This deviation between results obtained with the two techniques using the same model is a feature seen in all of our calculations and is also observed in others \cite{quasi}. We also note some difference in the parametric behaviours, particularly at low-$T$ and near $T_c$. This could be added to the growing body of evidence cautioning against using the RTA for precise quantitative studies of strongly-interacting systems.

\subsection{Shear viscosity}
\begin{figure}[!htbp]
	\includegraphics[width=\columnwidth]{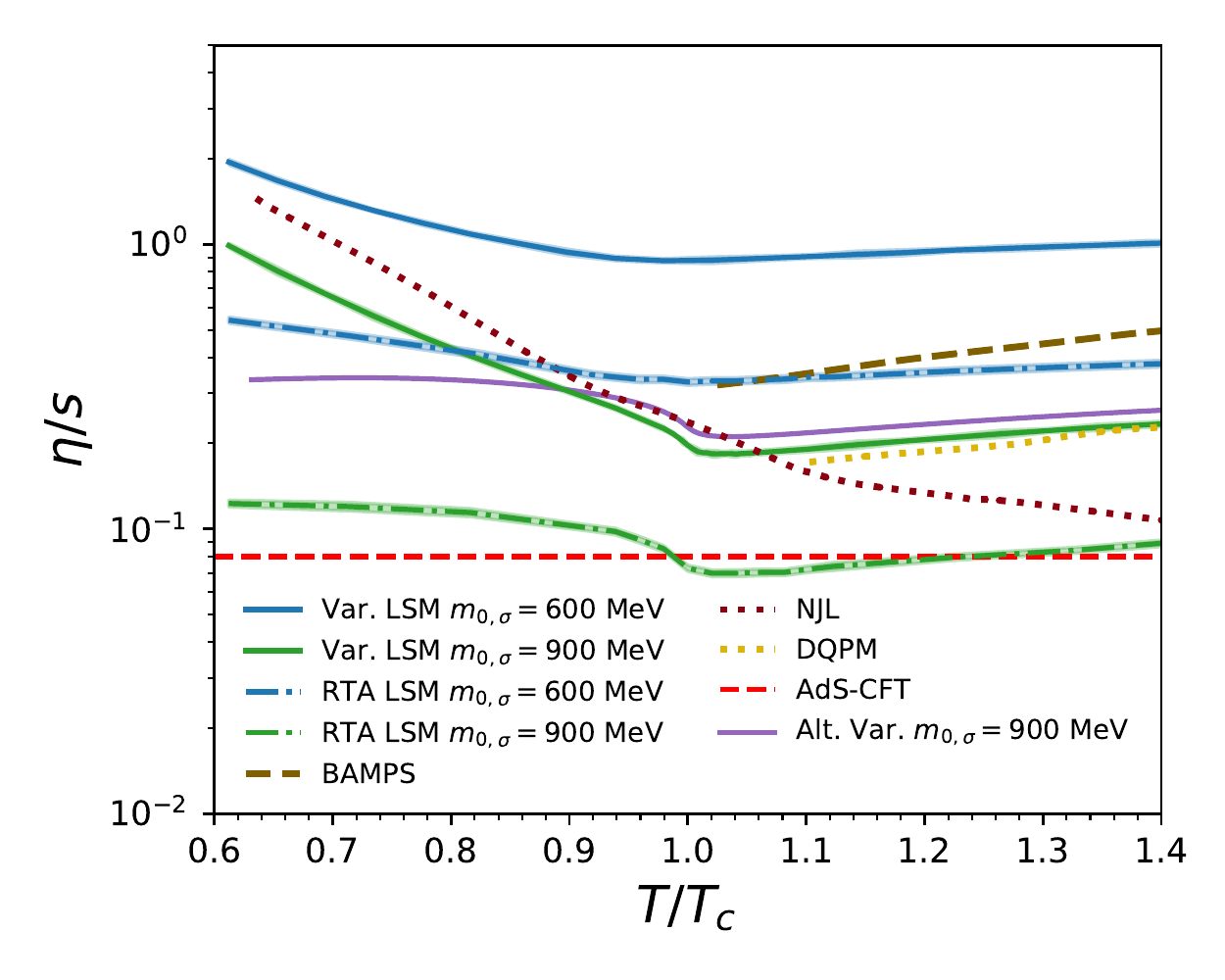}
	\caption{\label{fig:shearcomparison} The specific shear viscosity, $\eta/s$, is calculated in the relaxation time approximation and with the functional variation method, and compared to results using BAMPS \cite{BAMPS}, the dynamical quasi-particle model (DQPM) and NJL of \cite{njl-dqpm}, and AdS-CFT \cite{kss}. We additionally compare to the variational technique of \cite{quasi}, labelled as ``Alt. Var.".} 
	
\end{figure}

The shape and value of the shear viscosity to entropy density ratio for strongly-interacting matter is a topic of immense interest. Our results (Fig.~\ref{fig:shearcomparison}) again reveal that relaxation time calculations of the minimum value of $\eta/s$ can be as much as a factor of 3 lower than the value of a more precise calculation within the same theory. While the RTA calculation of $\eta/s$ with a vacuum $\sigma$ mass of 900 MeV approaches the KSS result of $1/4\pi$ \cite{kss}, the same calculation in the variational method does not. 

Many calculations of $\eta/s$ exist in the contemporary scientific literature and we again can not show an inclusive compendium here. It is appropriate to show a direct comparison to another similar calculation in the linear sigma model \cite{quasi}. As seen on Fig. \ref{fig:shearcomparison}, the numerical results reported here are close those seen in \cite{quasi}, with a difference increasing with decreasing temperatures. As $T$ shrinks to the lowest values explored here, an apparent plateau in $\eta/s$ is seen in \cite{quasi} -- and perhaps even a decrease -- whereas the values of specific shear viscosity calculated here follow an almost perfect exponential increase. In addition, the figure contains results obtained with BAMPS -- a relativistic Boltzmann equation solver \cite{BAMPS}, the Dual Quasi-Particle Model \cite{njl-dqpm}, and with the NJL model \cite{njl-dqpm}. These approaches again all yield results that differ over the range of temperatures chosen here. This is a recurrent theme and is consistent with the current state of affairs in the field. Several of those approaches use different degrees of freedom but parton-hadron duality for $T\sim T_c$ has the potential to minimize these differences. Finally, calculations of $\eta/s$ exist for lower values of $T$ \cite{Prakash:1993bt,Davesne:1995ms}. Given that the range of validity of those decreases with increasing temperature\footnote{The work in \cite{Prakash:1993bt} relies on chiral perturbation theory, for instance.} and our results have used the $s \to \infty$ limit, showing that they agree in some range of $T$ (they do) has questionable value. 

In most hydrodynamic applications, quantities are often fixed by the {\em ratios} of different transport coefficients. If the system can be characterized by a single relaxation time then, once we know one transport coefficient, others can be deduced by knowing these ratios. 
A natural question is that, while the use of the RTA is questionable for quantitative studies, perhaps the ratios between transport coefficients calculated using the RTA are close to those obtained using the variational technique. 
We explicitly considered the ratio of electrical conductivity to shear viscosity, $\sigma_{el}/\eta$. The value of the ratio using the RTA depends on choice of $m_{0,\sigma}$. Choosing the lower value more closely reproduces the ratio seen in the functional variation calculation. Even with the larger scalar-isoscalar masses, the ratios from the RTA are within $\sim 20$\% of those obtained using functional variations. The ratios obtained with both techniques are almost flat above $T_c$. We will show these results situated in context and in more detail in upcoming work.

\subsection{Bulk viscosity}
\begin{figure}[!htbp]
\includegraphics[width=\columnwidth]{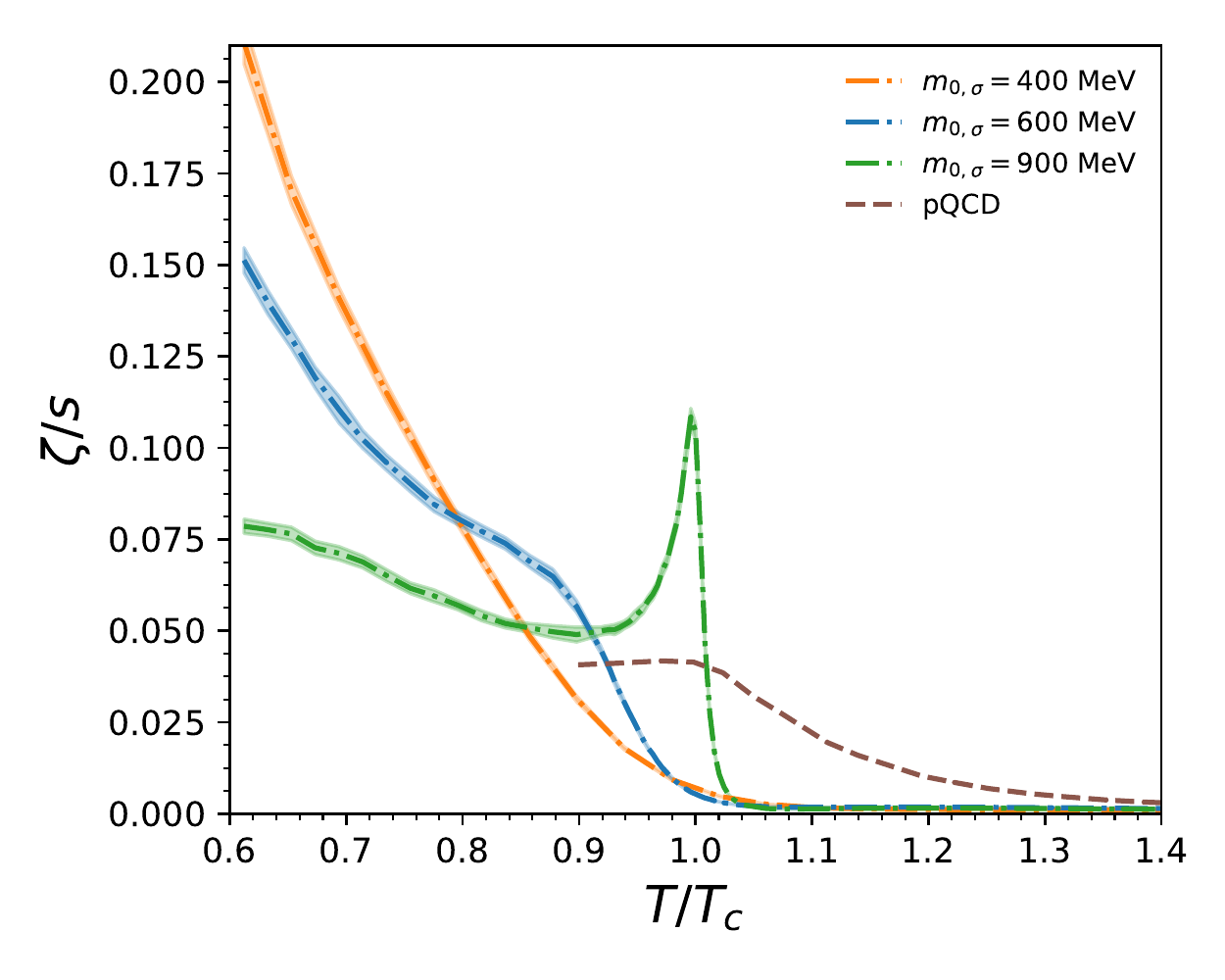}
\caption{\label{fig:RTbulk} The bulk viscosity to entropy density ratio for a variety of vacuum sigma masses in the relaxation time approximation. Comparisons are shown to pQCD \cite{Arnold:2006fz}. 
\label{bulk_viscosity}}
\end{figure}

We present the calculation of the linear sigma model bulk viscosity in the RTA in Fig.~\ref{fig:RTbulk}. The only computational advantage of the RTA presents itself here: zero modes of the collision matrix are not present. However, as we have established through the calculations of electrical conductivity and shear viscosity, we can produce only an estimate that exhibits the broad dynamics of the more precise calculation.

We do not calculate bulk viscosity in the variational technique because of zero modes, as discussed in Sec.~\ref{sec:variation}. We instead calculate the bulk viscosity in the relaxation time approximation as this approximation bypasses the issue.  We have shown in Figs.~{\ref{fig:sigmacomparison},~\ref{fig:shearcomparison}} evidence suggesting that the RTA is insufficiently precise for detailed studies, and discussions in the literature also suggest the RTA is insufficient for calculations of the bulk viscosity \cite{Czajka:2017wdo}. With those caveats in mind, one observes that the peak exhibited in Fig.~\ref{fig:RTbulk} by the LSM RTA calculation with the larger $\sigma$ mass is approximately a factor of 3 lower than that used in some hydrodynamics-driven phenomenological analyses \cite{Ryu:2015vwa,McDonald:2016vlt}. This must be understood in context: as made clear in Ref. \cite{ChunQM2019}, the bulk viscosity is currently not well constrained by systematic analyses of experimental data. 

As stated many times, the convergence of results for transport coefficients is currently not at hand. This is especially true for the bulk viscosity and it is clear that the proper exact treatment of bulk viscosity should be a priority for future investigations. One of the reasons for this emphasis is the special dual role enjoyed by $\zeta/s$. On one hand, in dynamical simulations the bulk viscosity reflects the resistance of the hydrodynamic system to volumetric deformation and therefore has a direct impact on the average transverse momentum of measured hadrons \cite{Ryu:2015vwa}. On the other hand, it can also be related to the nonconformality of the underlying theory, QCD \cite{Karsch:2007jc}.

\section{Conclusions}
In this work, we have used a comprehensive general framework for the calculation of transport coefficients in massive quasiparticle theories at finite temperature with inelastic processes. Using the linear sigma model, hadronic transport coefficients were calculated and compared to results from other theories. We have produced the first calculations of the electrical conductivity in the linear sigma model using the RTA and functional variational techniques. We have shown calculations for the shear viscosity in both methods, while a calculation of the bulk viscosity was only produced in the relaxation time approximation. Reasons for this choice were provided in detail. In all cases, we observe that the RTA and the variational results can differ by a factor of $\sim$3. This difference arises as a direct consequence of the RTA, in which the approximation becomes exact in the limit when only one species in the system is out of equilibrium at any given time. This is overly simplistic and has a large impact on the results. This should be interpreted as a cautionary flag for all calculations and should preface most -- if not all -- current theoretical attempts at a quantitative characterization of strongly-interacting matter out of equilibrium. This reinforces the need for precise calculations using realistic models of hadron and parton dynamics and for rigorous and systematic phenomenological extractions of transport coefficients from experimental relativistic heavy-ion data. This study should also provide impetus for further phenomenological applications by influencing,  and even providing, prior distributions in Bayesian analyses \cite{Paquet:2020rxl}.

\section{Acknowledgments}
We thank S. Hauksson for many useful discussions, P. Arnold for helpful correspondence, and G. Moore for an informative discussion on bulk viscosity. We thank P. Chakraborty and J. Kapusta for detailed and useful communications and S. A. Bass for a useful suggestion.
This work was supported in part by the Natural Sciences and Engineering Research Council of Canada (NSERC). Computations were made on the supercomputer Beluga from McGill University, managed by Calcul Qu\'{e}bec and Compute Canada. The operation of this supercomputer is funded by the Canada Foundation for Innovation (CFI), the minist\`{e}re de l'\'{E}conomie, de la science et de l'innovation du Qu\'{e}bec (MESI) and the Fonds de recherche du Qu\'{e}bec - Nature et technologies (FRQ-NT).

\appendix
\section{Mean field effects and Landau matching}
\label{appMF}
This section provides a summary of the Landau matching in \cite{MSc}, in turn based on that in  \cite{quasi}.

Using Eq.~(\ref{eq:phiregroup}), it is simple to rewrite Eq.~(\ref{eq:rewrite1}) as
\begin{eqnarray}
-\mathcal{U}\mathcal{A}^a + \mathcal{D}^\mn\mathcal{C}^a_\mn = 0.
\end{eqnarray}
and extract the functions $\mathcal{C}_{\mu\nu}^a$ and $\mathcal{A}^a$. 
Details of this are provided in \cite{MSc}.
\begin{eqnarray}
\mathcal{C}_{\mu\nu}^a &=& \frac{p^\mu p^\nu}{2TE_a} +\sum_{bcd} \label{eq:departureC} \frac{1}{1+\delta_{cd}}\int_{p_b,p_c,p_d} W(a,b|c,d)f_b^{eq} \nonumber\\
&\hspace{0.1in}& \hspace{0.5in}\times (C^d_\mn + C^c_\mn - C^a_\mn - C^b_\mn)
\end{eqnarray}
\begin{eqnarray}
\mathcal{A}^a
&=& \frac{1}{3TE_a} \left( (1- 3v_s^2 )(p^\alpha u_\alpha)^2 - m_a^2 \right)
\label{eq:mathcalA}\\
&+& \sum_{bcd} \int_{p_b,p_c,p_d} \frac{W(a,b|c,d)f_b^{eq}}{1+\delta_{cd}} \left(A_c + A_d - A_a - A_b \right) \nonumber
\end{eqnarray}
These ``departure functions'' characterize the shear and bulk departures from equilibrium.
An important subtlety is that in the departure function decomposition of the Boltzmann equation,
there is not a unique solution to $A^a$ in Eq.~(\ref{eq:mathcalA}). Infinite solutions can be generated by shifting a particular solution $A(E)$, e.g. 
\begin{eqnarray}
A'(E) = A(E)- a - bE
\end{eqnarray}
where $a$ is an arbitrary constant associated with particle conservation and $b$ is an arbitrary constant associated with energy conservation. This degree of freedom is related to the fact that the Boltzmann equation admits {\em summational invariants} \cite{cercignani2002relativistic}. We are restricting our scope to that with no chemical potential, thus all $a$ are $0$ as there are no particle conservation considerations. Thus, a particular solution $A_a'$ to the above can be related to all other solutions
\begin{eqnarray}
A'_a = A_a - b E_a.
\end{eqnarray}

Thus, by considering Eq.~(\ref{eq:deltatmn_dep})
and $u_\mu \Delta T^\mn = 0$, it is straightforward to conclude that
\begin{eqnarray}
0 &=& \sum_a \int_p E_a f_a^{eq} \left[A_a- bE_a\right].
\end{eqnarray}

Recalling the definitions of the single particle contributions to thermodynamic quantities,
it can be seen that
\begin{eqnarray}
b &=& \frac{1}{T^2 c_{V}}\sum_a \int_{p_a} E_a f_a^{eq} A_a.
\end{eqnarray}

We now consider the variational impact from the mean field effects on  Landau matching.
It is of particular importance to treat $\delta f$ correctly \cite{Czajka:2017wdo}. We take a small deviation from the equilibrium distribution function:
\begin{eqnarray}
f_a (x,p) = f_a^{eq} (E_{a,0}) + \delta f_a(x,p).
\end{eqnarray}
The single particle energy also takes an off-equilibrium shift
\begin{eqnarray}
E_a = E_{a,0} + \delta E_a.
\end{eqnarray}
If the equilibrium distribution function is expressed as a function of the true energy (including off-equilibrium shifts), then 
\begin{eqnarray}
f_a(x,p) = f_a^{eq}(E_a) + \delta \tilde{f}_a(x,p)
\end{eqnarray}
and
\begin{eqnarray}
\delta \tilde{f}_a (x,p) = \delta f_a(x,p) - \frac{\partial f_a^{eq}(E_a)}{\partial E_a} \delta E_a.
\end{eqnarray}
The shift in the energy density is therefore
\begin{eqnarray}
\Delta T^{00} &=& \sum_a \int_{p_a} \left(E_a \delta \tilde{f}_a (x,p) - \frac{f_a^{eq}(E_a)}{2T} \frac{d {m}_a^2}{dT}\delta T \right).\hspace{0.3in}
\end{eqnarray}
with
\begin{eqnarray}
\delta f_a &=& -e^{-E_a/T} \left(\frac{\delta E_a}{T} - \frac{E_a}{T^2}\delta T \right)\\
\therefore \delta \tilde{f}_a &=& e^{-E_a/T}\frac{E_a\delta T}{T^2}. \label{eq:ftilde}
\end{eqnarray}
We now return to the consideration of $\Delta T^{00}$.
\begin{eqnarray}
\Delta T^{00} &=& \sum_a \int_{p_a} \frac{1}{E_a}\left(E_a^2 - T^2 \frac{d {m}_a^2}{dT^2} \right) \delta \tilde{f}_a (x,p)\hspace{0.2in}\label{eq:deltaepsilon}
\end{eqnarray}
We obtain this final result by recalling $2TdT = dT^2$.
By definition, $u^2=1$. As a result, we \textit{satz} that $u^\mu u^\nu$ is the prefactor of the $T^2$ term \textit{an}. This follows from physical arguments, such as that this effective mass dependence does not have an impact on the pressure in the local rest frame. Thus, we generalize as follows:
\begin{eqnarray}
\Delta T^\mn = \sum_a \int_{p_a} \frac{1}{E_a}\left(p^\mu_a p^\nu_a - u^\mu u^\nu T^2 \frac{d {m}_a^2}{dT^2} \right) \delta \tilde{f}_a (x,p_a)\hspace{0.35in}\label{eq:deltatmunu}
\end{eqnarray}

Recall the definitions of shear and bulk viscosity
\begin{eqnarray}
\eta &=& \frac{2}{15} \sum_a \int \frac{d^3 p_a}{(2\pi)^3} \frac{|\mathbf{p}_a|^4}{E_a} f_a^{eq} C^a \\
\zeta &=& \frac{1}{3}\sum_a \int \frac{d^3 p_a}{(2\pi)^3} \frac{|\mathbf{p}|^2 }{E_a} f_a^{eq}A_a.
\end{eqnarray}

We now impose Landau matching with the new results on the effects of mean fields. This only modifies bulk viscosity $\zeta$.
As before, if a particular solution does not meet the Landau-Lifshitz matching condition, it can be made to comply by adding/subtracting a linear energy term
\begin{eqnarray}
\sum_a \int_{p_a} \frac{f^{eq}_a (x,p)}{E_a}\left[E_a^2 - T^2 \frac{d {m}_a^2}{dT^2} \right] \left[ A_a(E_a) - bE_a \right] = 0. \nonumber \\\label{eq:landaulifshitz}
\end{eqnarray}

We now use a relation to simplify the process and find the final result.
\begin{eqnarray}
\frac{dP}{dT} &=& \frac{dP}{d\epsilon}\frac{d\epsilon}{dT} = v_s^2 \frac{d\epsilon}{dT} 
\end{eqnarray}
Compiling Eqs.~(\ref{eq:ftilde},~\ref{eq:deltaepsilon},~\ref{eq:deltatmunu}) results in
\begin{eqnarray}
\Delta \epsilon = \Delta T^{00} &=&\frac{1}{T^2}\sum_a \int_p \left(E_a^2 - T^2 \frac{d {m}_a^2}{dT^2} \right)f^{eq}\delta T \label{eq:cvderivation} \hspace{0.15in}
\end{eqnarray}
\begin{eqnarray}
\Delta P = \Delta T^{ii} &=& \frac{1}{3T^2}\sum_a \int_p|\mathbf{p}|^2 f^{eq}\delta T.
\end{eqnarray}

Thus,
\begin{eqnarray}
0 &=& \frac{dP}{dT} - v_s^2 \frac{d\epsilon}{dT} \nonumber\\ 
&=& \sum_a \int_{p_a} f^{eq} \left[ |\mathbf{p}_a|^2 - 3 v_s^2 \left(E_a^2 - T^2 \frac{d {m}_a^2}{dT^2} \right) \right].\hspace{0.15in}
\end{eqnarray}
Using this expression, we may rearrange slightly and constrain the solution with the use of Landau matching. 
\begin{eqnarray}
\sum_a \int_{p_a} f^{eq} |\mathbf{p}|^2 &=& \sum_a \int_{p_a} f^{eq} \left[ 3 v_s^2 \left(E_a^2 - T^2 \frac{d {m}_a^2}{dT^2} \right) \right] \hspace{0.25in}\\
&=& 3 T^2 s
\end{eqnarray}
And as a result, it is possible to constrain $b$ in Eq.~(\ref{eq:landaulifshitz})
\begin{eqnarray}
0 &=& \sum_a \int_{p_a} \frac{1}{E_a}f^{eq}_a \left[E_a^2 - T^2 \frac{d {m}_a^2}{dT^2} \right]\times \left[ A_a - bE_a \right] \hspace{0.25in}
\end{eqnarray}
\begin{eqnarray}
b &=& \frac{v_s^2}{T^2 s} \sum_a \int_p \frac{1}{E_a}f^{eq}_a \left[E_a^2 - T^2 \frac{d {m}_a^2}{dT^2} \right] A_a
\end{eqnarray}
We may finally substitute $A_a'(E_a) = A_a - bE_a$ into the expression for bulk viscosity and we can conclude with an expression for bulk viscosity that meets the Landau-Lifshitz condition by construction
\begin{eqnarray}
\zeta&=&\frac{1}{3}\sum_a \int_{p_a} \frac{|\mathbf{p}_a|^2 }{E_a} A_a f_a^{eq}\left( |\mathbf{p}|^2 - 3v_s^2 \left[E_a^2 - T^2 \frac{d {m}_a^2}{dT^2} \right]\right)\nonumber \\
\end{eqnarray}
Substituting the solution for $A_a$ from the relaxation time approximation yields Eq. (\ref{eq:bulkmeanfield}).

\section{Details of the functional minimization}
\label{appFuncMin}
Explicitly, the collision term of the functional (Eq.~(\ref{eq:Q})) is 
\begin{eqnarray}
&&(\chi_\ij , \mathcal{C}\chi_\ij) \nonumber\\
&& \hspace{0.1in}= \frac{\beta^3}{8}\sum_{abcd} \int_{p_a,p_b,p_c,p_d} W(a,b|c,d)f_0^a f_0^b\nonumber\\
&&\hspace{0.2in}\times\big[\chi_\ij^a(p_a)+\chi_\ij^b(p_b)-\chi_\ij^c(p_c)-\chi_\ij^d(p_d) \big]^2\nonumber\\
\end{eqnarray}
and the source is
\begin{eqnarray}
\left(\chi_\ij,S_\ij\right) &=& -\beta^2 \sum_a \int_p f_a^{eq}q^a\chi^a.
\end{eqnarray}

We now encounter the first point where the masses enter the theory: in the use of the convenient delta function expansion
\begin{eqnarray}
&&\delta^{(0)}(E_a+E_b-E_c-E_d)\nonumber\\
&& \hspace{0.25in}= \int_{-\infty}^{\infty} dw \delta(w+E_a-E_c)\delta(w-E_b+E_d).
\end{eqnarray}
where we recognize $w$ as the energy transfer. We additionally define a momentum transfer $\mathbf{q}$ such that $p_c = p_a + \mathbf{q}$, $p_d = p_b -\mathbf{q}$. 
To remove the integration over cosines, we must expand these delta functions and derive limits.
\begin{eqnarray}
&&\delta (w-E_b + E_d) \nonumber\\
&&\hspace{0.25in}= \frac{E_d}{p_b q}\delta\left(\cos\theta_{p_b q} - \frac{2wE_b-t+m_d^2-m_b^2}{2p_b q}\right)\label{eq:ebtheta}
\end{eqnarray}
\begin{eqnarray}
\cos\theta_{p_a q} &=& \frac{m_a^2+t+2wE_a-m_c^2}{2p_a q}
\end{eqnarray}
and
\begin{eqnarray}
&&\delta (w+E_a - E_c) \nonumber\\ 
&& \hspace{0.25in}= \frac{E_c}{p_a q}\delta\left(\cos\theta_{p_a q} -\frac{m_a^2+t+2wE_a-m_c^2}{2p_a q} \right). \label{eq:eatheta}
\end{eqnarray}
Implicit in each of these final delta functions is a theta function that ensures that energy is conserved; $\Theta(E_b - w)$ in Eq.~(\ref{eq:ebtheta}) and $\Theta(E_a + w)$ in Eq.~(\ref{eq:eatheta}).

The masses do not modify the Jacobians, so these are the same for the massless case.
Performing the cosine integrals trivially using the delta function will yield limits to ensure that the delta functions are satisfied. In order to find the limits, we solve the inequality 

\begin{eqnarray}
\cos^2\theta_{p_a q} = \left(\frac{m_a^2+t+2wE_a-m_c^2}{2p_a q}\right)^2 \leq 1.
\end{eqnarray}
If we consider only elastic processes, we can make the further simplification that $m_a = m_c$ and $m_b = m_d$. Adding the assumptions that $E_{p_a,p_b} \geq 0$, $m_{p_a,p_b} = m_{p_c,p_d} \geq 0$ yields the following bounds:

\begin{eqnarray}
E_a &\geq& \frac{1}{2} \left( \sqrt{\frac{q^2(4m_a^2 + q^2 - w^2)}{q^2-w^2}}-w\right)
\end{eqnarray}
\begin{eqnarray}
E_b &\geq& \frac{1}{2} \left( \sqrt{\frac{q^2(4m_b^2 + q^2 - w^2)}{q^2-w^2}}+w\right)\\
|w| & < & q
\end{eqnarray}

Our calculations also include inelastic collisions, but the limits for inelastic processes are significantly more involved and are excluded due to space constraints. However, the phase space can be thoroughly explored by taking the clear kinematic limits and checking numerically to see that the conditions from the delta functions are satisfied. We include details here to clarify what differences arise without taking the small momentum transfer approximation in a massive case. The biggest consideration is in the treatment of angular cross-terms, which are radically different.
Returning to the integral,

\begin{widetext}
\begin{eqnarray}
(\chi_\ij , \mathcal{C}\chi_\ij) &=& \frac{\beta^3}{(4\pi)^6}\sum_{abcd} \int_0^\infty p_a^2 dp_a q^2dqk^2 dp_b\int_{-q}^{q} dw \int_0^{2\pi}d\phi_q\frac{E_c E_d}{p_a p_b q^2E_a E_b E_cE_d}|\mathcal{M}|^2 f_0^a f_0^b \label{eq:collisiontilde} \\
&&\hspace{0.5in}\times\left[\chi_\ij^a(p_a)+\chi_\ij^b(p_b)-\chi_\ij^c(p_c)-\chi_\ij^d(p_d) \right]^2\Theta(E_a + w)\Theta(E_b-w)\nonumber
\end{eqnarray}
\end{widetext}

$\chi(p)$ can then be expanded in a basis,
\begin{eqnarray}
\chi(p) &=& \sum_{m=1}^{N} a_m \phi^{(m)}(p)\label{eq:chi}
\end{eqnarray}
where we successfully use the same basis as that in \cite{Arnold:2000dr},
\begin{eqnarray}
\phi^{(m)}(p) &=& \frac{(p/T)^m}{(1+p/T)^{N-1}}
\end{eqnarray}
where $N$ is the size of the basis, which has been chosen such that it converges quickly. The coefficients $a_m$ are maximized in order to maximize the functional $Q$, which we will denote as $Q_{max}$.
We must now also repeat the treatment of delta functions for cosines to account for masses. All our results for limits and angular factors recover those of \cite{Arnold:2000dr} when $m\rightarrow 0$.
\begin{eqnarray}
\cos\theta_{p_a q} &=& \frac{2wE_a +t + m_a^2 - m_c^2}{2p_a q} \\
\cos\theta_{p_b q} &=& \frac{2wE_b - m_b^2+m_d^2 - t}{2p_b q}\\
\cos\theta_{p_c q} &=& \frac{2wE_c -t + m_a^2 - m_c^2}{2p_c q}\\
\cos\theta_{p_d q} &=& \frac{-2wE_d - m_b^2+m_d^2 + t}{2p_d q}\\
\cos\theta_{p_a p_c } &=& \frac{2E_a E_c+t-m_a^2-m_c^2}{2p_a p_c}
\end{eqnarray}
\begin{eqnarray}
\cos\theta_{p_b p_d} &=& \frac{2E_b E_{d}+t-m_b^2-m_d^2}{2p_b p_d}\\
\cos\theta_{p_a p_b} &=& \cos\theta_{p_a q}\cos\theta_{p_b q}+\sin\theta_{p_a q}\sin\theta_{p_b q}\cos\phi \hspace{0.35in}\\
\cos\theta_{p_a p_d} &=& \cos\theta_{p_a q}\cos\theta_{p_d q}+\sin\theta_{p_a q}\sin\theta_{p_d q}\cos\phi\\
\cos\theta_{p_c p_b} &=& \cos\theta_{p_c q}\cos\theta_{p_b q}+\sin\theta_{p_c q}\sin\theta_{p_b q}\cos\phi \\
\cos\theta_{p_c p_d} &=& \cos\theta_{p_c q}\cos\theta_{p_d q}+\sin\theta_{p_c q}\sin\theta_{p_d q}\cos\phi
\end{eqnarray}

What remains is to properly deal with the $\chi$ factor. Knowing that the matrix elements in the LSM are constants has the potential to simplify the forms of some of the $\phi$ integral, but this is properly done at the end of the manipulations.

In not taking the small momentum transfer approximation, we must consider all of the angular factors. As a result, we require proper treatment of angular to account for the inelastic processes that take place in the linear sigma model.

\begin{eqnarray}
\left[\chi_\ij^a(p_a)+\chi_\ij^b(p_b)-\chi_\ij^c(p_c )-\chi_\ij^d(p_d) \right]^2
\end{eqnarray}
This produces terms such as 
\begin{eqnarray}
2\chi_\ij^a(p_a)\chi_\ij^b(p_b) &=& 2\chi^a(p_a)\chi^b(p_b)P_l(\cos\theta_{p_a p_b}). \hspace{0.25in}
\end{eqnarray}
Taking the analogy to $\phi$ as shown before, the right hand side becomes
\begin{eqnarray}
\left(\phi^a_m(p_a)\phi^b_n(p_b) + \phi^a_n(p_a)\phi^b_m(p_b)\right)P_l(\cos\theta_{p_a p_b}).\hspace{0.2in}
\end{eqnarray}
This occurs for all the cross terms and angles between particles. The result of this computation is inserted into Eq.~(\ref{eq:collisiontilde}), but is not included here due to length. Care must be taken with angular cross-terms in Eq.~(\ref{eq:collisiontilde}). These are explicitly 
\begin{eqnarray}
I_\ij (\hat{p}_a) I_\ij (\hat{p}_b) = P_l(\cos\theta_{p_ap_b})
\end{eqnarray}
where $P_l(\cos\theta)$ is the $l^{th}$ Legendre polynomial and refers to the spin of the underlying transport coefficient. As discussed in Sec.~\ref{sec:variation}, the appropriate values are 0 for the bulk viscosity (a scalar), 1 for the conductivity (a vector), and 2 for the shear viscosity (a tensor). 

The electrical conductivity, at least in the case of the LSM, is a simple case as the source for neutral particles is explicitly 0 and we expect $\delta f$ to be identical only for particles with identical mass and reactions. This makes it possible to easily identify components of the collision matrix that do not contribute. The shear and bulk viscosity will require appropriate consideration of different $\delta f$ for $\pi$ and $\sigma$ mesons. 

Most importantly, in shear and bulk viscosity, we must consider a cross-coupling between the sigma and the pi with appropriate consideration for their different departures from equilibrium. This means that instead of an $N \times N$ collision matrix, we instead have a $2N \times 2N$ collision matrix. Note that it is not $4N\times 4N$ because for the viscosities, the pions are identical. We therefore sum the three pions in the $\pi$ segment of the above matrix and the $\pi\sigma$ interactions in the $\pi\sigma$ sections above. In the expansion in Eq.~(\ref{eq:chi}),
one species will have expansion coefficients $a_m$, but the other will then have the expansion coefficient $a_{N+m}$. 
We now turn to a discussion of the source.

The assembly of the collision matrix $\tilde{C}$ and the source vector $\tilde{S}$ require some details to ensure that interactions are properly implemented. The assembly of these into block components most clearly demonstrates this process. $\tilde{C}$ is assembled is by calculating each component with only one species $C^\sigma$, $C^\pi$ and then the interactions between particle species, $C^{\sigma\pi}$, and then compiling into blocks within the respective matrices.
\begin{eqnarray}
\begin{bmatrix}
C^{\sigma} & C^{\sigma\pi}\\
C^{\sigma\pi} & C^{\pi}
\end{bmatrix}
\end{eqnarray}
The notation below is somewhat simplified, $C^\pi$ can be further expanded into \[C^\pi =\begin{bmatrix} C^{\pi^0} & C^{\pi^0\pi^+} & C^{\pi^0\pi^-}\\
C^{\pi^+\pi^0} & C^{\pi^+} & C^{\pi^+\pi^-}\\
C^{\pi^-} & C^{\pi^-\pi^+} & C^{\pi^-\pi^0}\\
\end{bmatrix}\]
although this becomes more cumbersome to show in the full collision matrix. $C^\sigma$ contains all reactions with only $\sigma$ and the off diagonals contain pion-sigma reactions.
Similarly, the source is comprised of sub-components that are compiled into
\begin{eqnarray}
\begin{bmatrix}
S^\sigma \\
S^\pi
\end{bmatrix}
\end{eqnarray}
The term $S^\pi$ may be decomposed in the same manner as $C^\pi$.

Thus, $Q_{max}$ becomes
\begin{eqnarray}
Q_{max} = \frac{1}{2} \begin{bmatrix} S^\sigma & S^\pi \end{bmatrix}
\begin{bmatrix}
C^{\sigma} & C^{\sigma\pi}\\
C^{\sigma\pi} & C^{\pi}
\end{bmatrix}^{-1}
\begin{bmatrix} S^\sigma \\
S^\pi \end{bmatrix}.
\end{eqnarray}

\bibliography{transportcoefficients.bib}

\end{document}